\documentclass[
        final,
        inline,
        ]
        {ieee}
\usepackage{times}
\usepackage{tablefootnote}
\usepackage{color}
\usepackage{amsmath,amssymb}
\usepackage{bm}% bold math
\usepackage{graphicx}
\usepackage{subcaption}
\usepackage{caption}
\begin{document}

%----------------------------------------------------------------------
% Title Information, Abstract and Keywords
%----------------------------------------------------------------------
\title[Enge Detector]{%
       The Focal plane Detector Package on the TUNL Split-pole Spectrograph}

% %format author this way for journal articles.
\author[Marshall]{
      Caleb Marshall
      \authorinfo{%
      C. Marshall is with the Department of Physics,
      North Carolina State University, Raleigh NC, 27695, USA
      }
    \and
      Kiana Setoodehnia
      \authorinfo{%
      K. Setoodehnia is with the Department of Physics,
      North Carolina State University, Raleigh NC, 27695, USA}
    \and
      Katie Kowal
      \authorinfo{%
      K. Kowal is with IDA Science and Technology Policy Institute,
      1899 Pennsylvania Ave. NW, Washington DC, 20006, USA}     
    \and
      Federico Portillo
      \authorinfo{%
      F. Portillo is with the Department of Physics,
      North Carolina State University, Raleigh NC, 27695, USA} 
    \and
      Arthur E. Champagne
      \authorinfo{%
      A. E. Champagne is with the Department of Physics,
      The University of North Carolina at Chapel Hill, Chapel Hill NC, 27514, USA
      } 
    \and
      Stephen Hale
      \authorinfo{%
      S. Hale is with Integrated Science Support, Inc., Smithville, MO  64089}
    \and
      Andrew Dummer
      \authorinfo{%
      A. Dummer is with Polarean, Inc., Durham, NC 27713}
    \and
      and Richard Longland
      \authorinfo{%
      R. Longland is with the Department of Physics,
      North Carolina State University, Raleigh NC, 27695, USA}
}
% specifiy the journal name
%\journal{IEEE Transactions on Instrumentation and Measurement, 2018}
% Or, when the paper is a preprint, try this...
\journal{IEEE Transactions on Instrumentation and Measurement, \copyright 2018 IEEE, Accepted For Publication.}

% Or, specify the conference place and date.
%\confplacedate{Ottawa, Canada, May 19--21, 1997}

% make the title
\maketitle               

% do the abstract
\begin{abstract}
  A focal plane detector for the Enge Split-pole Spectrograph at
  Triangle Universities Nuclear Laboratory has been designed. The detector package
  consists of two position sensitive gas avalanche counters, a gas proportionality
  energy loss section, and a residual energy scintillator. This setup allows both
  particle identification and focal plane reconstruction. In this paper we will
  detail the construction of each section along with their accompanying electronics and data acquisition.
  Effects of energy loss throughout the detector, ray tracing procedures, and resolution as a function of fill pressure and bias voltage are also investigated.
  A measurement of the $^{27}\!$Al$(d,p)$ reaction is used to demonstrate detector performance and to illustrate a Bayesian method of energy calibration.
\end{abstract}

% do the keywords

% start the main text ...
%----------------------------------------------------------------------
% SECTION I: Introduction
%----------------------------------------------------------------------
\section{Introduction}

\PARstart Nuclei heavier than beryllium are mostly created through nuclear reactions occurring in stellar interiors.   
Furthermore, for elements with $A{<}70$ nucleosynthesis proceeds largely through the nuclear capture of charged particles \cite{starstuff}.
However, at astrophysical energies, the coulomb barrier heavily suppresses the reaction cross section and inhibits the stellar reaction rate. Therefore, direct measurement of astrophysically important reactions at the relevant energies is difficult, and in some cases impossible.

If a direct measurement of the reaction of interest is not feasible, the reaction rate can still be estimated by using indirect methods. Indirect methods aim to improve knowledge of a reaction by measuring energies,
angular momenta, spectroscopic factors, asymptotic normalization coefficients, or other properties of the relevant nuclear states \cite{iliadis}.
Examples of indirect measurements include studies of transfer, stripping, and charge exchange reactions with both stable and radioactive beams \cite{stable} \cite{RIB}.
Many of these reactions produce outgoing charged particles, whose energy is determined by the state that was populated in the residual nucleus. Detecting these
particles with high energy resolution is a key requirement of any indirect study, where precise knowledge of excitation energies is important and high
level densities can potentially obscure states of interest. 

Magnetic spectroscopy is one method for achieving high energy resolution \cite{enge}.
Magnetic spectrographs utilize magnetic fields to spatially separate
particles according to their energy and charge to mass ratio. A charged particle moving with velocity $\mathbf{v}$
through a magnetic field $\mathbf{B}$ is subjected to a force given by:

\begin{equation}
  \label{eq:lorentz}
  \mathbf{F} = q \mathbf{v} \times \mathbf{B}.
\end{equation}

\noindent In the case of magnetic spectrographs, a perpendicular force is applied, and the path through the magnetic field can be described {(non-relativistically)} by:

\begin{equation}
  \label{eq:rig}
  B \rho = \frac{mv}{q} = \frac{\sqrt{2mE}}{q},
\end{equation}

\noindent where $m$ is the mass of the particle, $v$ is its velocity, $q$ is its charge, $E$ is the kinetic energy of the particle,
$B$ is the magnetic field of the spectrograph, and $\rho$
is the radius of the particle's circular orbit through the spectrograph. The product $B\rho$ is known as the
magnetic rigidity.

\begin{figure}
  \centering
  \includegraphics[width=.5\textwidth]{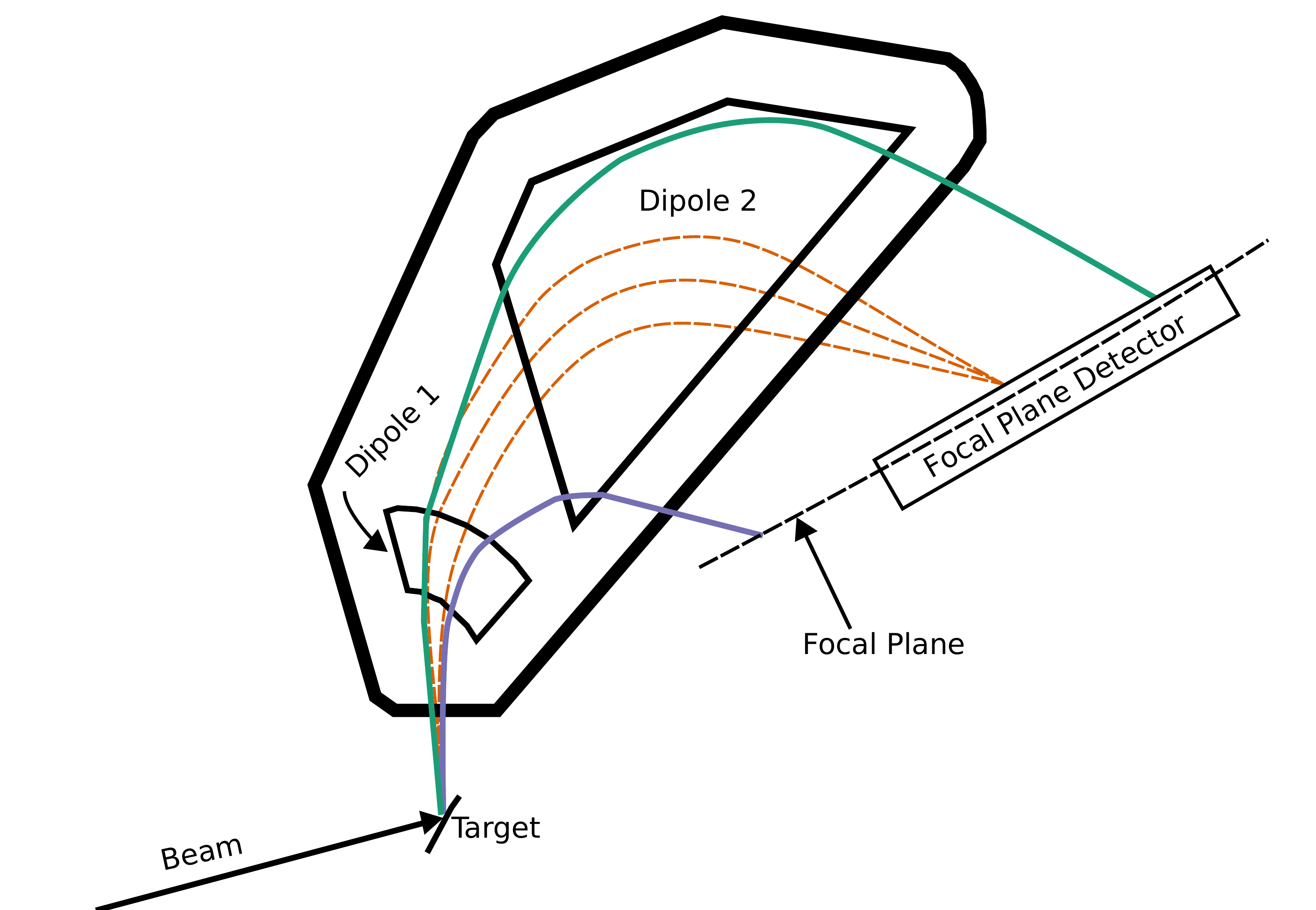}
  \caption{{(color online) Example particle trajectories through a Split-pole spectrograph. The spectrograph focuses (orange dashed lines)
      particles with similar magnetic rigidities onto a slightly curved focal plane. The detector that is described in this work sits on
      this focal plane and measures the relative positions of these particles.}}
    \label{fig:engeoptics}
\end{figure}

The spectrograph at the Triangle Universities
Nuclear Laboratory (TUNL) is an Enge Split-pole \cite{splitpole}, so-called because a single sector magnet is
split into two in order to provide second-order double focusing with additional vertical
focusing \cite{enge}. After leaving the second dipole, charged particles with similar magnetic rigidities will converge to a point.
For the Split-pole, the locus of focal points for different magnetic rigidities forms a dispersive image of the target along a gently curved focal plane that lies at a $41.5^{\circ}$ angle to the magnetic exit. A focal plane detector positioned along this plane will record the positions and, therefore, the magnetic rigidities of these particles.
This setup is represented pictorially in Figure \ref{fig:engeoptics}.

Along with a FN Tandem Van de Graaff accelerator, this spectrograph forms the foundation of a modern facility devoted entirely to experimental nuclear astrophysics \cite{cosmos}. This facility is capable of delivering a variety of light ions to target with a millimeter spot size.
This beam spot size, combined with the horizontal magnification of the Split-pole spectrograph, $M_x \! \sim \! 0.34$, implies a peak of width of $\sim\!0.34$ mm on the focal plane.  
Therefore, this facility requires a detector specifically designed to detect high energy, low mass particles with sub-millimeter spatial resolution.
In this paper an updated focal plane detector system will be presented, which is modeled after the
design outlined in Ref. \cite{hale}. 

\begin{figure}
  \centering
  \includegraphics[scale=.2]{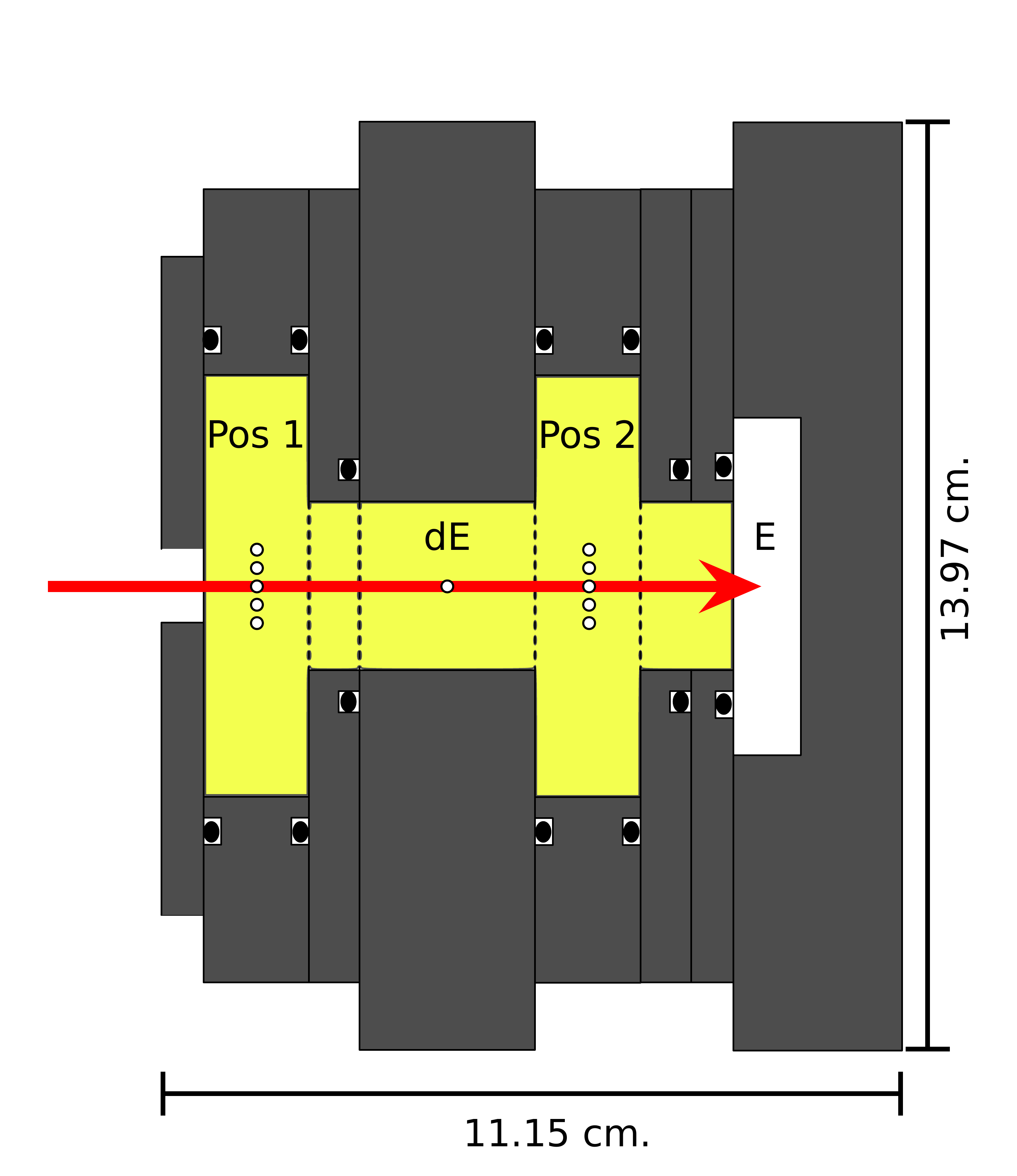}
  \caption{(color online) A cross section view of the whole detector package. The red arrow represents the direction of incident particles and
    the black ovals show the location of the o-ring seals. {The approximate location of the anode wires throughout the detector are indicated by the white circles.
    The gas filled regions are indicated by light yellow shading. Though not indicated in the figure, the length of the detector is $71.12$ cm.}}
  \label{fig:wholedet}
\end{figure}

The focal plane detector we describe here is an
assembly of two position sensitive
avalanche counters, a
gas proportionality counter ($\Delta E$ section), and a residual
energy scintillator ($E$ section).
The $\Delta E/E$ detector combination is used to distinguish
between the different species of light ions. The inclusion of a second position section
allows particle paths to be reconstructed (see Section III.C). These paths can be
used to optimize peak resolution offline. A cross section of the entire detector can be seen in Figure \ref{fig:wholedet}.

The fabrication of each detector section will be discussed in Sec.II. 
Sec.III will be devoted to characterizations, including: energy straggling simulations,
the effects of kinematic broadening, and experimental tests of optimal operating parameters.
Finally, a Bayesian method for energy calibration will be presented in Sec.IV, which 
was used in the analysis of $^{27}\!$Al$(d,p)$ to confirm sub-millimeter position resolution and
to infer energy levels of $^{28}\!$Al.

\section{Design Characteristics and Detector Fabrication}

\subsection{Position Section}

\begin{figure}
  \centering
  \includegraphics[scale=.3,angle=90]{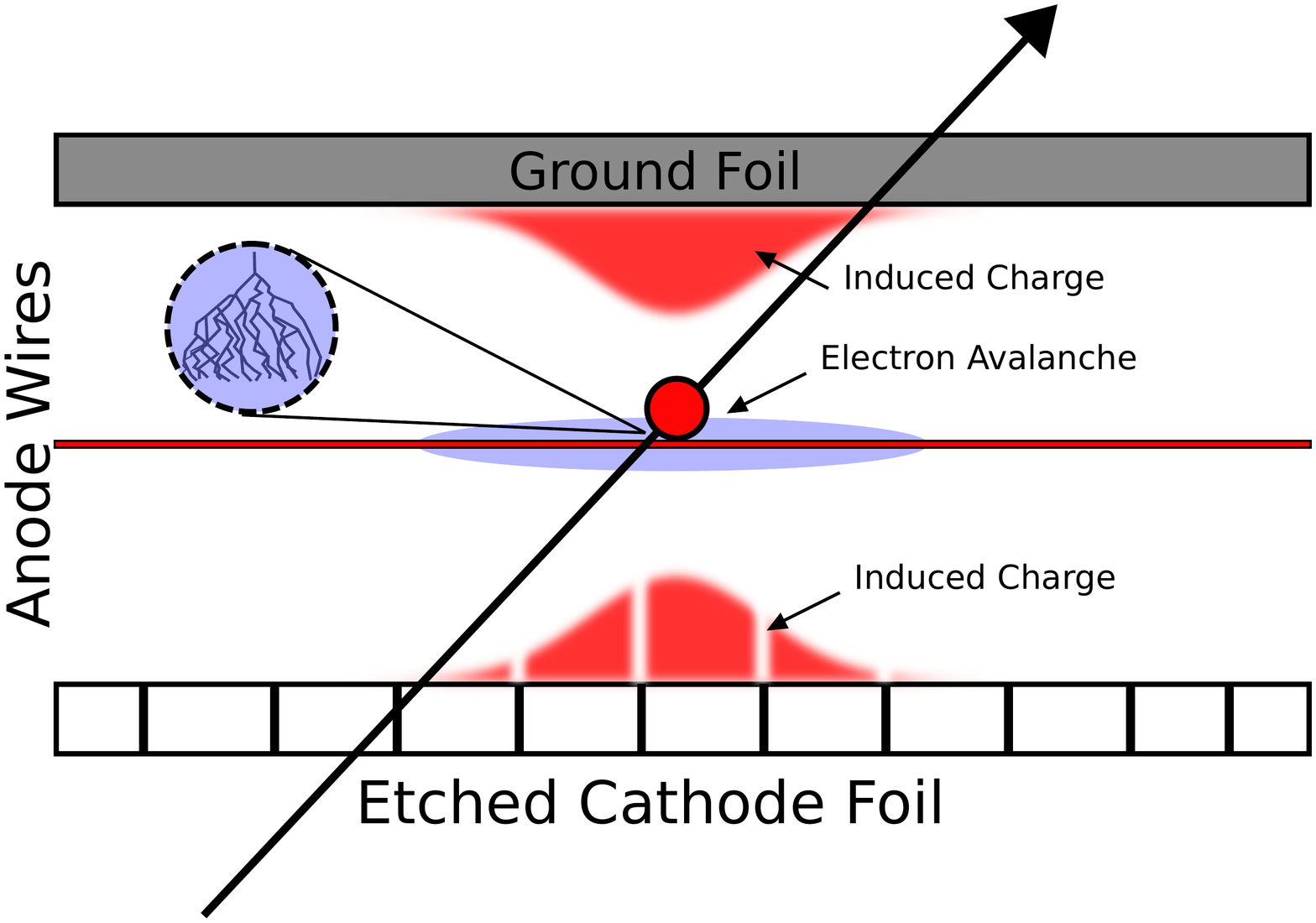}
  \caption{(color online) Cartoon of the principles of operation. When a charged particle enters the detector, ionization occurs on the fill gas, and an induced charge is created on the etched {foil} and grounded {foil}.}
  \label{fig:ion}
\end{figure}

Position measurements of particles leaving the high magnetic field region of the Split-pole spectrograph are performed by two position sensitive avalanche counters. The positions are measured near the entrance of the detector and again before the particles are stopped in the total energy scintillator. The position-sensitive avalanche counters operate as follows, and are
represented pictorially in Fig.\,\ref{fig:ion}. 5 high voltage anode wires are located within each counter between two cathode foils made of aluminized Mylar. {These counters sit inside the detector chassis that is pressurized to 200 Torr with circulating isobutane. The pressurized environment is isolated from the high vacuum of the spectrograph with a $12.7\textnormal{-} \mu m$ thick Kapton entrance window. After the particles pass through the window and begin to travel through the isobutane, the gas is ionized.} If the ionization events occur within the electric field of the counters, electrons will be rapidly accelerated towards the anode wires setting off a series of secondary ionization events creating an electron avalanche \cite{knoll}. The avalanche is negatively charged and localized around the particle's position as it passes the anode. Negative charge induces a positive charge on both the cathode foils. The foil closer to the entrance of the detector is electroetched \cite{etch}, and is described in detail below. Etching creates electrically isolated strips that are connected together via a delay line. Thus, as charge is carried out of the detector from each strip, it is exposed to a different amount of delay. The timing difference from each side of the detector corresponds to the measurement of position. If the charge was only distributed over one strip our position resolution would be limited by the strip width. However, distributing the charge over multiple strips allows an interpolation of the composite signal, thereby improving the spatial resolution to the sub-millimeter level. As the particle exits the position sections, it passes through the grounded cathode foil that helps shape and isolate the electric field from the anodes.  

{Position sensitive avalanche counters are commonly designed to have
pickup strips parallel to the incident particle path \cite{msu}
\cite{heavy} \cite{parikh} \cite{argonne_det}; however, the etched foils in our detector sit
perpendicular to the particle path. This type of setup has also been
used in the focal plane detector for the Q3D spectrograph at the Maier-Leibnitz Laboratory
\cite{vert} and the now decommissioned Q3D at Brookhaven National Lab \cite{BNLDetector}.
These designs have been shown to have excellent position resolution.
Additionally, if cathode foils are used, the number of wires required can be drastically reduced; thus, easing maintenance of the
detector. However, these designs are not suited towards heavy ion reactions, where
the cathode foils would provide additional scattering surfaces that degrade mass resolution.
The effects of these foils on the energy loss of light particles are explicitly examined
in Section III.A, and are found to have a negligible impact on the resolution. }

Below we discuss the methods used to fabricate the position section assemblies with particular attention paid towards the etched cathode planes, delay line, and timing electronics.

\subsubsection{ElectroEtching Technique}

{The design of our position section is critically reliant on having precisely etched cathode foils. These foils should have evenly spaced, electrically isolated
strips, which necessitates a process to remove the aluminum coating from the Mylar foil. Electric discharge etching was chosen to create the cathode foils.}
It was shown in Ref. \cite{etch} that this technique produces clean, uniform lines. Chemical etching with sodium hydroxide has also been shown to work, for example see Ref. \cite{vert}, but difficulties arise with the precise application of sodium hydroxide and the cleaning of the reaction products. We also found that electroetching could reliably produce etched foils in less than a day, which reduces time and effort required for routine maintenance. Each strip is $2.54\textnormal{-} mm$ wide, with each etched line being $0.03\textnormal{-} mm$ wide. The strips are etched on $0.3\textnormal{-}\mu m$ thick single sided aluminized Mylar.

\begin{figure}
  \centering
  \includegraphics[scale=.7]{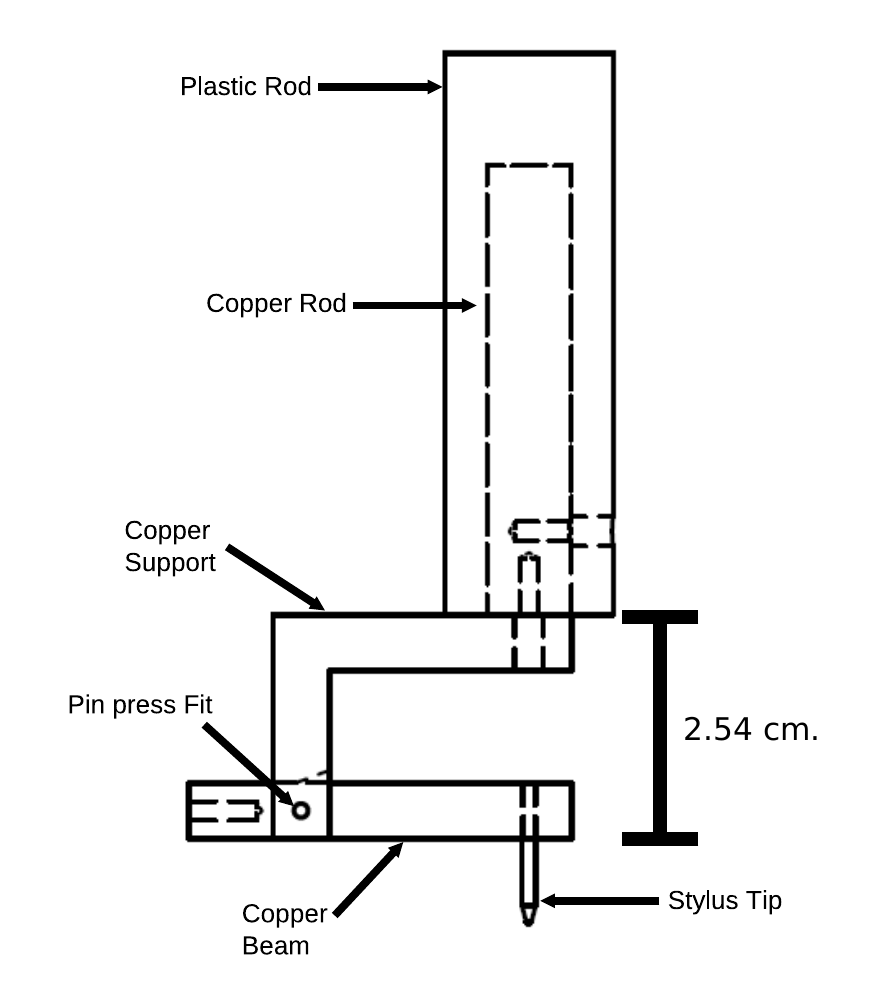}
  \caption{Drawing of the etching apparatus. The biased stylus tip is allowed to follow the curvature of the Mylar thanks to the pivoting copper arm.
  The plastic housing insulates the milling machine from the biasing potential. The dashed lines indicate threaded holes for screws.}
  \label{fig:etch}
\end{figure}

Our particular setup consists of a tungsten tipped stylus attached to a copper assembly pictured in Fig. \ref{fig:etch}. {The etching is performed using a milling machine programmed with G-code. To isolate the copper rod from the spindle of the machine, a plastic covering rod was used.} During the machining process, it is of vital importance to keep
good electrical contact between the Mylar and stylus tip. {To ensure this condition several steps were taken.} The stylus arm was attached to its base with a pivot. This design allows the tip to follow the natural curvature of the Mylar. {Additionally, the Mylar is carefully clamped to the milling table with two $5.08\, \textnormal{cm} \times 5.08\, \textnormal{cm}$  grounded metallic plates. These plates were found to provide the proper grounding throughout the etching process. It was also found that periodic sanding of the tungsten tip is necessary to prevent aluminum buildup.} The tip itself was held at $-15\ V$ during the process. This voltage was found to produce clean lines while reducing the possibility of damaging sparks. Reference \cite{etch} found that negative polarity produced cleaner lines when examined under an electron microscope.
       
\subsubsection{Delay Lines}

The delay line consists of 20 delay chips {with 10 taps per chip. Each tap provides $5$ ns of delay making the total delay across the line $1\, \mu$s.}
Copper plated G-10 boards where machined to align copper strips with the etched pickup strips. This creates electrical contact between the etched pickup strips and the delay chip legs. The legs are attached via pin inserts on the back of the G-10 boards. The chips, Data Delay Devices 1507-50A \cite{chips}, have a 50 $\Omega$ impedance, which matches that of the signal cables.
Weldable BNC feedthroughs attached to NPT threads provide a vacuum tight method for connection to the delay line signal cables. It must also be noted that the error in delay per tap is quite high at $\pm 1.5\ ns$, which could lead to non-linearity in the delay to position conversion \cite{msu}.
{Following the suggestions in Ref. \cite{widths}, this effect is minimized by ensuring the ratio of the cathode strip width (2.54 mm) and the distance between the anode and the cathode (3.00 mm) is around $0.8$.}    

\subsubsection{Position Section Assembly}

\begin{figure}
  \centering
  \includegraphics[scale=.30]{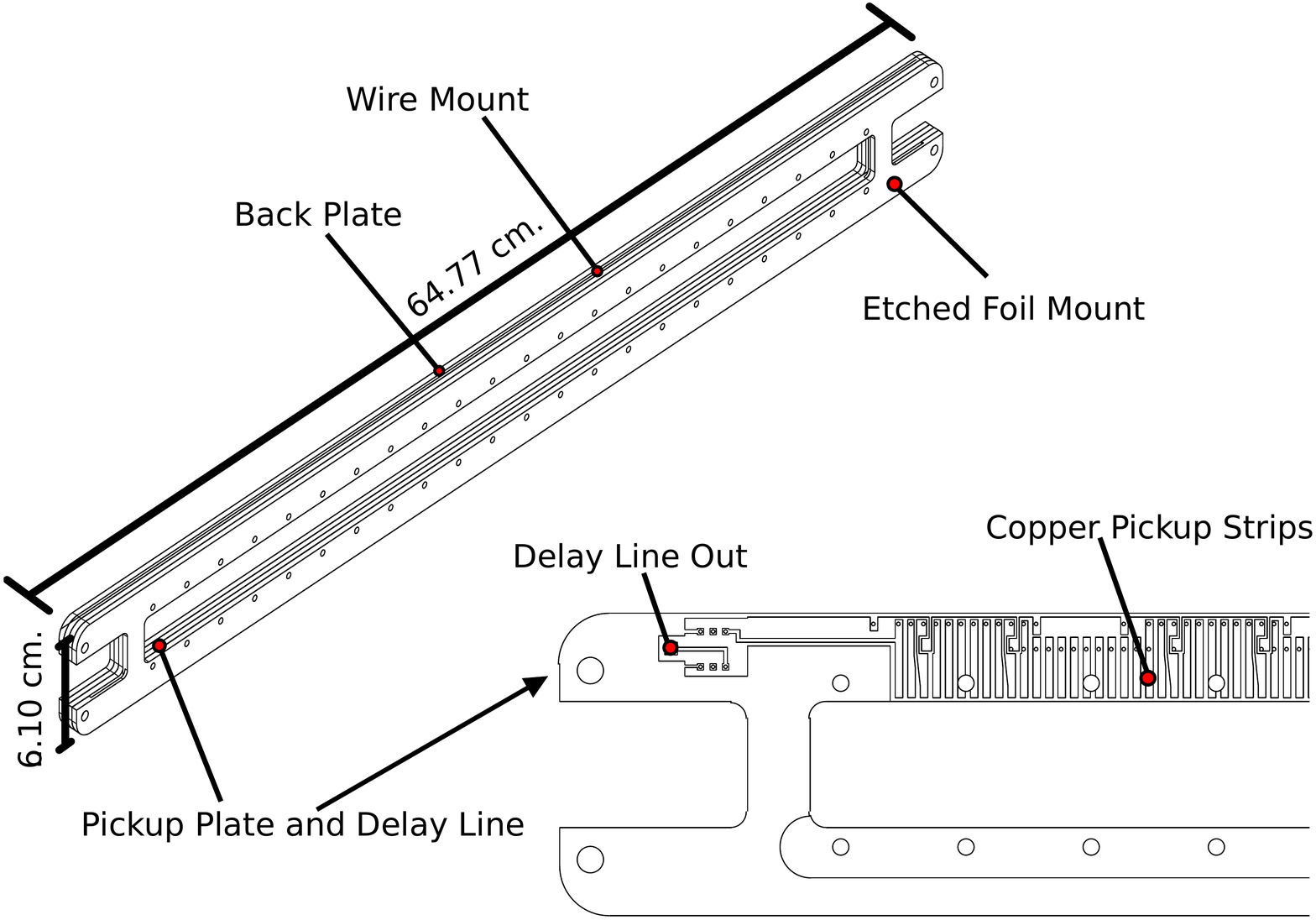}
  \caption{A model of the position section assemblies. From front to back we have: the copper plated front plate to which the etched cathode foil is taped, the copper
    plated G-10 board with pickup strips and mounts for the delay line chips, the anode wire plane board, and the back board to which the grounded plane is taped.
    The expanded region shows the copper pickup strips that make contact with the cathode foil. The delay line is attached to the back via pin inserts that are
    at the top of each strip.}
  \label{fig:pos}
\end{figure}

The delay line, cathode {foil}, anode wires, and grounded {foil} are all housed in the position section assembly shown in Fig. \ref{fig:pos}. Four metallic screws bring the copper plated top into electrical contact with the detector body, which is grounded. Plastic screws ensure proper contact between the cathode {foil} and the delay line while maintaining electrical isolation with the rest of the board.
 
Five gold plated tungsten wires $25\textnormal{-} \mu m$ in diameter are used for the anodes. {The wire spacing is $4 \, mm$.} They are surrounded by the cathode {foils}. These {foils} are made of $0.3\textnormal{-}\mu m$ thick aluminized Mylar, either single sided for the etched cathode foils, or double sided for the grounded cathode {which were purchased from Goodfellow Cambridge Ltd.} The Mylar sheets are secured to both the detector and position assemblies using double sided tape. Care is taken to ensure good electrical connection between the detector and the grounding foils because the tape is an insulator. 

Accurate measurement of the charged particles position requires a well localized electron avalanche. This requirement means that we must operate the position sections at higher voltages than would be required of a proportional counter \cite{knoll}. In order to prevent sparking and allow voltages of ${\sim} 2000\ V$, insulating acrylic coating is applied to high risk areas and $1\ \textnormal{N}$ of tension is applied to the wires. The tension is necessary to keep the wires straight, which is required for a uniform electric field and to reduce sparking.
Isobutane was chosen as the fill gas, following the suggestions of Ref. \cite{gas}.

\subsubsection{Position Section Electronics}
Hereafter signals will be referred to based on whether they exit the detector on the side corresponding to a high value of the magnetic rigidity (high energy) or a low one (low energy).
Inside the focal plane chamber, each of the four position signals are sent through fast timing preamplifiers. After preamplification, the signals are processed through an Ortec 863 quad Timing Filter Amplifier (TFA). The final shaping and noise rejection before our timing analysis is provided by a Constant Fraction Discriminator (CFD). Thresholds on each of the channels are adjusted to match the output levels of the TFA, which are on the order of $~300 $mV. After the final signal shaping, the signals from the high energy end of the detector are used to start an Ortec 567 Time to Amplitude Converter (TAC), while the low energy signals are all subject to a $1\ \mu \textnormal{s}$ delay and used to stop the TAC. This delay ensures the stop signal always occurs after a start signal for a real event. The output from the TAC is sent into a CAEN V785 peak sensing Analog-to-Digital Converter (ADC), so that it can be recorded and later analyzed.

\subsection{$\Delta E $ Section}
\subsubsection{$\Delta E $ Assembly}

The $\Delta E$ section of the detector is a gas proportional counter, which consists of a single $12.7\textnormal{-} \mu m$ diameter anode wire and two grounded cathode planes.
The front cathode plane is the other side of the front position section's grounded cathode plane. The back plane is another strip of aluminized Mylar.
This Mylar plane is taped directly onto the detector body and checked for proper electrical contact. Due to low breaking tension of the anode wire, it is held taut by hand, and soldered onto NPT threaded feedthroughs. The wire is biased to $1000\ V$ to ensure that the charge collection is proportional to the energy loss of the particle.     

\subsubsection{$\Delta E$ Electronics}

The $\Delta E$ section of the detector's signal is processed with an in-house charge sensitive preamplifier based on the Cremat CR-110 operational amplifier, which provides a $1\ \mu s$ shaping time \cite{cremat}. After the preamplifier, an Ortec $572$A amplifier is used to shape the signal before it is sent to the ADC.  

\subsection{Residual Energy Section}

\subsubsection{Paddle Scintillator}

Particles are stopped, and residual energy deposited, in a Saint-Gobain BC-404 organic plastic scintillator. The BC-404 is sensitive to $\alpha$ and $\beta$ radiation, and is recommended for fast timing \cite{gobain}. The timing response makes it an ideal trigger for the current data acquisition system and planned $\gamma$-ray coincidence measurements. The dimensions are $28.25''$ long by $2''$ wide by $.25''$ thick. These dimensions are customized to cover the length of the detector and ensure all light particles will stop within the active volume. In order to maximize the amount of light collected along the entire length of the scintillator, it is wrapped in thin, reflective aluminum foil and Tyvek. Reference \cite{wrapping} demonstrated that Tyvek has an increased light output compared to the aluminum wrapping; however, it was unable to
hold pressure, so the aluminum foil was also used to create a sealing surface.

\subsubsection{Optical Fibers}

Early iterations of the $E$ section used a light guide to couple the paddle scintillator
to the Photomultiplier Tube (PMT); however, this design added significant weight and length to
the detector. To avoid the rigid constraints of light guides, optical fibers were chosen to gather and transmit light to the PMT.

The fibers are $1\textnormal{-mm}$ diameter Bicron BCF-91A, which shift the wavelength of the violet/blue scintillated light ($380-495\ \textnormal{nm}$) of the BC-404 into the green spectrum ($495-570\ \textnormal{nm}$) \cite{fibers}.  
Following the suggestions of Ref. \cite{wrapping}, the optical fibers were spaced $5\ \textnormal{mm}$ apart
to maximize light collection. Eight $1\-\textnormal{mm}$ deep grooves were machined in the scintillator.
The fibers were placed into these grooves, and secured with BC-600 optical cement.
{A light tight tube is used to bend the fibers to the PMT that sits on the top of the detector.} 

\subsubsection{Photomultiplier Tube}
Matching the emitted light of the light fibers while maintaining a compact package were the main requirements for the PMT. 
The Hamamatsu H6524 has a spectral response of $300-650\ \textnormal{nm}$, a peak sensitivity of $420\ \textnormal{nm}$, and
a quantum efficiency of $27\%$ \cite{hamamatsu}. These features provide the highest quantum efficiency available for the wavelengths of interest.
The 10-stage dynode structure provides a gain of $1.7 \times 10^6$ with an anode bias of $-1500$ V.
Although the detector is located outside the Split-pole's high magnetic field region, a magnetic shield was incorporated into the tube assembly to prevent possible interference.

\subsection{{$E$ Electronics and Event Structure}}

The dynode signal from the PMT is split to provide both timing and energy information. {Energy signals are processed through
an Ortec $572$A amplifier and then recorded.} Timing signals go through a TFA and CFD to generate an {event count}.
A count from the $E$ detector triggers the master gate for the data acquisition system {, which can be vetoed if the ADC buffer is full.} 
If a trigger is not vetoed, a $10\textnormal{-} \mu s$ gate is generated {, and the ADC records all
coincident signals.}
Additionally, a Time-to-Digital Converter (TDC) is used to further restrict coincidence requirements in software.
{Using the $E$ signal to generate the ADC gate, as opposed to the position sections, avoids a position dependent gate timing.}
{Count rates} are recorded for all detector signals, gates generated, and gates vetoed due by ADC busy signal. This setup allows us to easily diagnose electronic problems and adjust beam current to keep the acquisition dead time low ($<\!10\%$).

\section{Detector Characterization}

\subsection{Geant4 Simulations}

\begin{figure}
  \centering
  \includegraphics[width=0.4\textwidth]{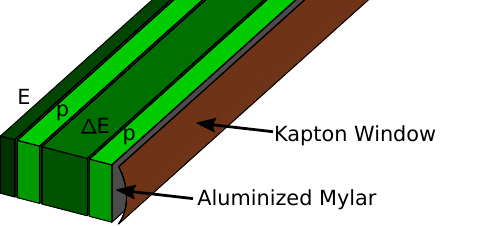}
  \caption{(color online) Geant4 model of the detector system. The bowed Kapton
    window is represented as a copper color here. The entire volume
    between that window and the total energy scintillator is filled
    with isobutane gas at 200 Torr and divided by aluminized Mylar to
    account for straggling in the detector. Sensitive volumes are
    defined to track particles through the active detector components,
    and are shown in various shades of green. "p" refers to the position sections.}
  \label{fig:Geant4Model}
\end{figure}

Simulations of the detector have been developed using
Geant4.10.03~\cite{Geant4} to investigate the effect of geometry and materials on
detector performance. Particles are expected to scatter as they pass
through the detector, thus degrading the detector resolution. Possible
scattering elements are the Kapton entrance window, aluminized Mylar cathode
{foils}, and isobutane gas. The Geant4 model consists of the Kapton window,
curved to simulate bowing from the pressure differential under
operation, isobutane gas volume, all cathode foils, and BC-404
scintillator. The isobutane gas volume is defined in the model to fill the entire
volume between this Kapton window and the scintillator. Finally, an
aluminum foil is included in front of the scintillation detector. The
Geant4 model is shown in Fig.~\ref{fig:Geant4Model}.

To investigate the maximum possible {{position}} resolution
of the detector (i.e., that which does not include electronic noise or
position resolution determined by the finite pickup pad size), we
model a beam of monoenergetic particles focused on the center of the
front position section. These particles impinge the front face of the
detector at {{$43.5^{\circ} \pm 9.5^{\circ}$}}. The incident
angle for each simulated particle is calculated to focus them
horizontally in the center of the front position section.

\begin{figure*}
  \centering
  \includegraphics[width=\textwidth]{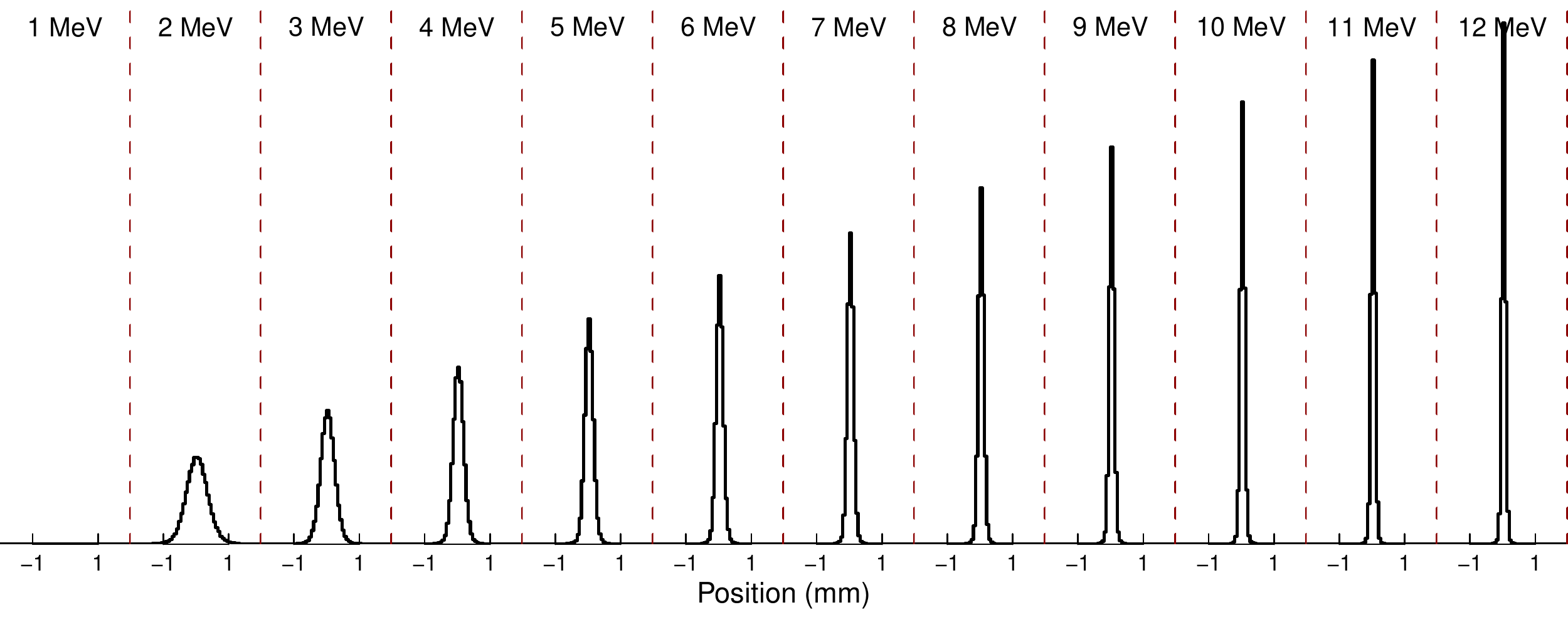} 
  \caption{(color online) {{Simulated}} position
    {measurements} for a range of incident proton energies. Shown are
    histograms of the {{simulated}} position (with respect
    to the focal point at 0 mm) at incident energies between 1 MeV and
    12 MeV. Each incident beam energy run is separated by a dashed,
    red line. Note that for 1-MeV protons, no ``detection'' is made
    because the particles stop before reaching the position
    section. At low energies, multiple scattering from the gas and
    Kapton window dominates, which causes the observed peak
    broadening.}
  \label{fig:FrontPos-protons}
\end{figure*}

The {{simulated particle tracks as they pass through the
    detector}} are recorded and saved in root files~\cite{ROOT}. Post
processing is then performed for particle tracking, position
determination, and energy deposition
calculations. {{Simulated position measurements are
    determined from where the particles pass the center of the front
    position section.}} In the absence of scattering, these positions
are expected to be identical in the horizontal direction. However,
once scattering in the gas, aluminized Mylar cathodes, Kapton, and
aluminum foils is included, simulated positions are expected to
exhibit some widening. This is indeed the case, as shown in
Fig. \ref{fig:FrontPos-protons}. At low proton energies, the
{{simulated position measurement}} is spread due to scattering of the
particles in the gas and foils. Indeed, 1-MeV protons are stopped in
the gas and do not traverse the front position section, thus their
positions are not recorded. At higher energies, though, straggling effects
are found to be minimal.

\begin{figure}
  \centering
  \includegraphics[]{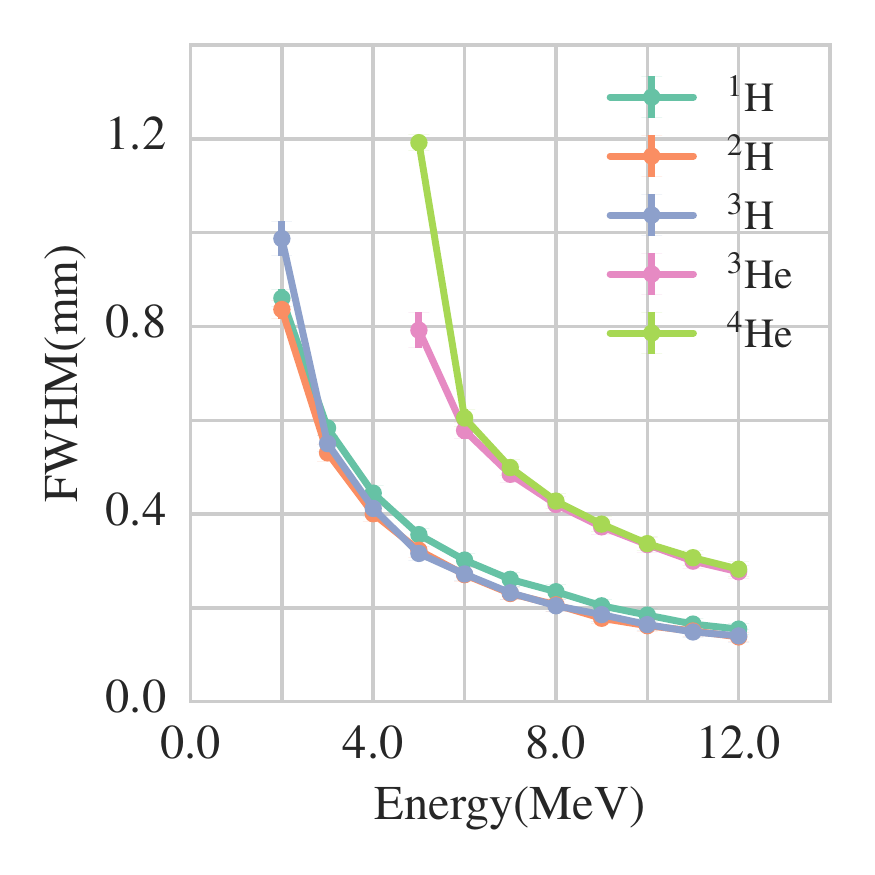}
  \caption{(color online) {{Simulated}} detector resolution
    for a range of beam particles of energies 1 MeV to 12
    MeV. Clearly, this detector design is not feasible for low energy
    helium isotopes, {{although a thinner Kapton entrance
        foil could be considered}}. Note that these results do not
    reflect sources of broadening other than scattering off
    detector elements.}
  \label{fig:Resolution}
\end{figure}

The resolution, taken as the Full-Width at Half-Maximum (FWHM) of the peaks
in Fig.~\ref{fig:FrontPos-protons} was determined for a range of
particles: protons, deuterons, tritons, $^3$He, and $^4$He. The
results of these simulations are shown in
Fig.~\ref{fig:Resolution}. Energy cut-offs are exhibited at about 2
MeV for hydrogen isotopes, and 6 MeV for helium isotopes. This finding
allows us to develop experiments using the detector accounting for
these limitations.

Though there is precedence for placing the cathodes in the particle
path, e.g. Ref. \cite{vert}, it is desirable to quantify their effects
on detector resolution by scattering.  Within the Geant4 model,
elements were selectively removed from the detector to find their
effects on the resolution. 7-MeV $\alpha$-particles were used to
perform this test, given that they should be strongly affected by
scattering. With all elements in place, a clear Gaussian position
distribution is found, with a maximum resolution of 0.48(2)~mm.  After
removing the aluminized Mylar foils, the resolution is unchanged
within statistical fluctuations. The isobutane was also removed from
the simulations to determine the effect of scattering in the gas and
found only a small effect, improving the resolution to 0.47 (2)~mm.
Finally, if the Kapton window is removed in the simulation, the
detector resolution becomes 0.06 (2)~mm. Clearly, the Kapton window
limits the simulated detector resolution. However, some vacuum
interface is necessary, so this element cannot be easily replaced.
The results of the simulations show that the effects of energy
straggling through the cathode foils and gas are not a limiting factor
on detector resolution.

\subsection{Kinematic Broadening Corrections}

{The location and shape of the Split-pole's focal plane changes based on the reaction kinematics. The finite acceptance solid angle of the spectrograph means
  the angular dependence of the outgoing particle's momentum will greatly degrade the resolution if no corrective action is taken \cite{enge}\cite{ionoptics}\cite{matrix}.
  For the Split-pole this is typically done by moving the focal plane detector to the new focal plane location \cite{enge}.
  To demonstrate the theory behind this correction, the ion optics of the spectrograph are examined using a transfer matrix  \cite{matrix}\cite{partA}.}

{The first order effects of kinematic broadening can be expressed using a phase space that consists of the entrance angle $\theta_i$, the exit angle $\theta_f$, the beam image $x_i$, the focal plane image $x_f$, and the momentum spread $\delta$, which is defined as:}

\begin{equation}
  \label{eq:7}
  \delta = \frac{\Delta p}{p} .
\end{equation}

Conservation of momentum implies that $\delta$ will be constant for a given trajectory. The coordinate system assumed here sets the beam direction as $+z$, beam left as $+x$, and up as $+y$. The full optics of the Split-pole would require an expanded phase space, but these extra terms have no impact on the first order angular correction. The final coordinates can be expressed in terms of the initial ones using the first order transfer matrix elements. For $x_f$ these are:

\begin{equation}
  \label{eq:9}
  x_f = (x_f|x_i)x_i + (x_f|\theta_i)\theta_i + (x_f|\delta) \delta ,
\end{equation}

{The design of the spectrograph is such that $(x_f|\theta_i) = 0$, which means that $x_f$ has no angular dependence with a constant $\delta$. First order kinematic broadening can be introduced by Taylor expanding $\delta$ around $\theta_i$ giving:}

\begin{equation}
  \label{eq:8}
  \delta(\theta_i) = \delta_0+\frac{\partial \delta}{\partial \theta_i}\theta_i = \delta_0 - K \theta_i ,
\end{equation}

\noindent where the kinematic factor $K$ is defined as the change in the momentum shift with a change in angle. Care must be taken with the sign of $K$ as it is dependent on both the laboratory setup and reaction kinematics. For the setup of our spectrograph, the reaction angle is with respect to beam left; thus, $\theta_i$ is positive, and we use normal kinematics, meaning a lower momentum with increased angle, leading to a negative sign on $K$.

{Correcting for this effect amounts to removing the dependence of $x_f$ on $\theta_i$. This can be done by displacing the detector in the $z$ direction. A change in $z$
  relates $\theta_f$ to $x_f$ and introduces two additional $\theta_i$ terms, which can be used to compensate for the kinematic shift. The expression for $x_f$ becomes:}

\begin{multline}
  \label{eq:12}
  x_f = (x_f|x_i)x_i + (x_f|\delta) \delta_0-K(x_f|\delta)\theta_i + \Delta z (\theta_f|\theta_i)\theta_i + \\
  \Delta z (\theta_f|x_i) x_i + \Delta z (\theta_f|\delta) \delta_0 - K \Delta z (\theta_f|\delta) \theta_i
\end{multline}

\noindent{Setting the $\theta_i$ terms equal to zero and solving for $\Delta z$ gives: }

\begin{equation}
  \label{eq:10}
  \Delta z = \frac{KDM}{1-KM(\theta_f|\delta)} \approx KDM ,
\end{equation}

\noindent where the quantities $M = (x_f|x_i)=\frac{1}{(\theta_f|\theta_i)}$ called the magnification and $D = (x_f|\delta)$ called the dispersion have been introduced. The approximation is valid when $K$ is relatively small, so that the denominator is close to unity. While $M$ and $D$ can be calculated theoretically, we chose to find an empirical fit between $K$ and $\Delta z$ in order to ensure maximum resolution.

$K$ can be determined for a given reaction using energy and momentum conservation, which produces the formula \cite{enge}:

\begin{equation}
  \label{eq:2}
  K = \frac{(M_bM_eE_b/E_e)^{1/2} \sin \theta}{M_e+M_r-(M_bM_eE_b/E_e)^{1/2} \cos \theta} ,
\end{equation}

\noindent where e references the ejected particle, b is for beam, and r is for the residual particle. 

\begin{figure}
  \centering
  \includegraphics[]{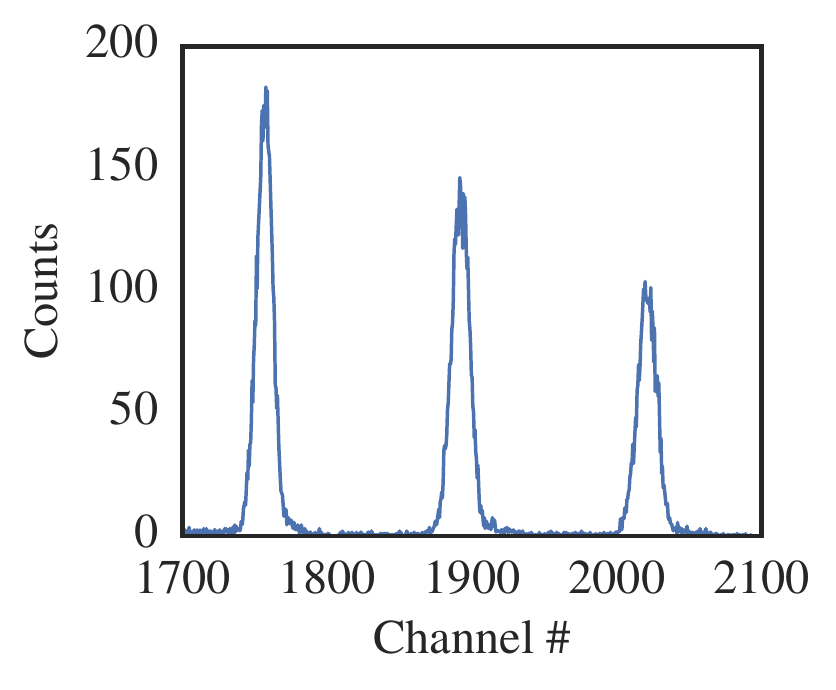}
  \caption{Example proton spectrum when detector is off focal plane with 3-slit aperture. The reaction is $^{12}$C$+$p elastic scattering at $\theta_{\textnormal{Lab}}=20^{\circ}$ and $E_{Lab} = 12$ MeV.
  The different peak intensities reflect the rapid variance of the cross section with the detection angle.}
  \label{fig:peaks}
\end{figure}

\begin{figure}
  \centering
  \includegraphics[]{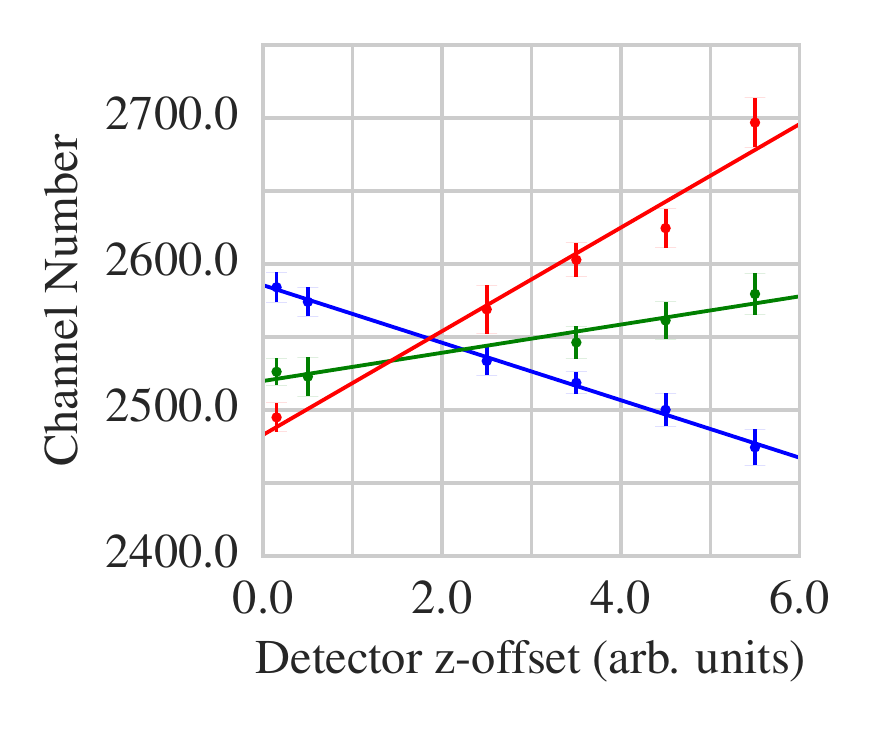}
  \caption{(color online) An example of measured peak centroids and the linear fit with respect to the $z$ position of the detector read in terms of a stepper motor voltage. Data is from the elastic scattering of protons off $^{197}\!$Au at $\theta_{\textnormal{Lab}} = 30^{\circ}$ and $E_{Lab} = 12$ MeV.}
  \label{fig:au}
\end{figure}

One possible method for finding an optimal $z$ for a given $K$ is described in Ref. \cite{parikh}. Using this method, the optimal $z$ position is found by moving the detector through the focal chamber and minimizing the width of a chosen peak; however, this method does not give much feedback during the run, as peak width can be hard to determine without careful peak fitting. Instead, a three-slit aperture was used, and the detector was displaced along its $z$ range. This aperture serves to discretize the acceptance solid angle into three narrow ranges of $\theta$. When the detector is off the focal plane, three particle groups will be observed as shown in Fig. \ref{fig:peaks}. When the detector is on the focal plane these groups should converge; thus, the detector is swept across the depth of the focal plane chamber and a linear fit of the accompanying peak positions is found, as shown in Fig. \ref{fig:au}.

Fig. \ref{fig:au} also demonstrates that while the effects of kinematic broadening can be limited by moving the detector position, they cannot be corrected entirely. This is due to higher order optical effects and uncertainties in the peak positions. Since the precise location of the convergence point of the lines is not measured, the $z$ value is taken to be the average of the vertices of the triangle formed by the line intersections. Discretizing the acceptance aperture of the spectrograph was found to be an effective method for producing a linear
calibration of the detector position in order to minimize resolution loss.

\subsection{Ray Tracing}

\begin{figure}
  \centering
  \includegraphics[width=.5
  \textwidth]{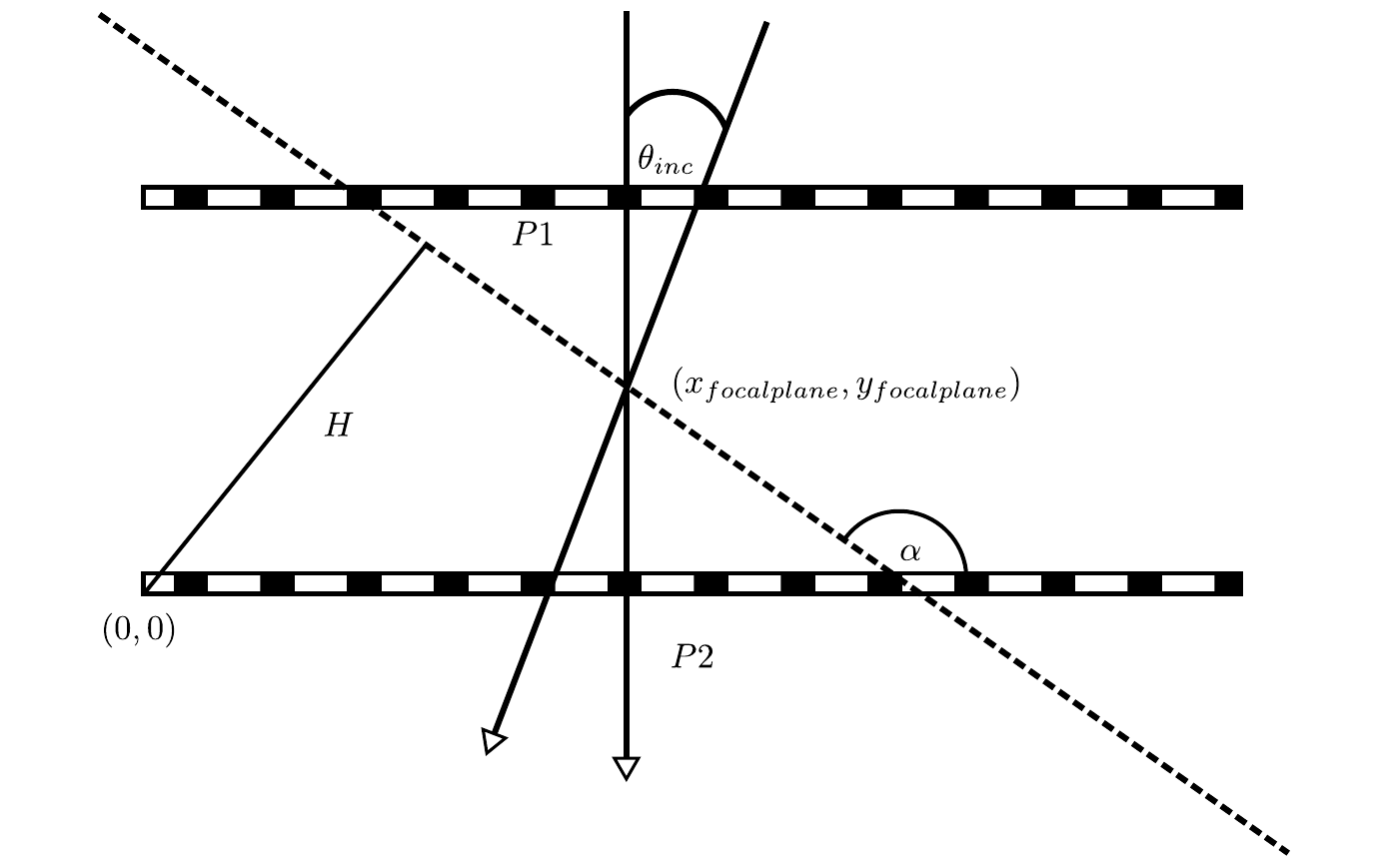}
  \caption{Geometrical schematic for ray tracing. The
  dashed line represents the focal plane, while the striped lines represent the detectors position sections. $H$ is the distance of the focal
    plane from the selected origin, $\alpha$ the angle between the back position and the focal plane, $P1$, $P2$, and $(x_{focalplane},y_{focalplane})$ give the coordinates of the particles at the front,back and focal plane respectively, and $\theta_{inc}$ is the incident angle relative to the ray that corresponds to the center of the reaction angle distribution.}
  \label{fig:ray}
\end{figure}

Even with extensive calibrations, the resolution of the detector is negatively affected with even small deviations from the focal plane \cite{raytrace} \cite{ugalde}. These effects can be remedied using ray tracing procedures using the two position measurements. Following the geometry of \cite{raytrace} (pictured in Fig. \ref{fig:ray}), we can write the observed positions according to:

\begin{equation}
  \label{eq:3}
  \begin{aligned}
    x_{focalplane} = \frac{\frac{H(P1-P2)}{\cos \alpha}+SP2}{S+\tan \alpha(P1-P2)},\\
    y_{focalplane} = \frac{S(\frac{H}{\sin \alpha}-P2)}{S \cot \alpha  + (P1-P2)},    
  \end{aligned}
\end{equation}

\noindent where, following Fig.\ref{fig:ray}, $P1$ is the observed position on the front position section, $P2$ is the position at the back position section, $\alpha$ is the angle between the focal plane and the detector, $H$ is the distance from the back edge of the detector to the focal plane, and $S$ is the distance between the front and back position sections.

\begin{figure}
  \centering
  \includegraphics[]{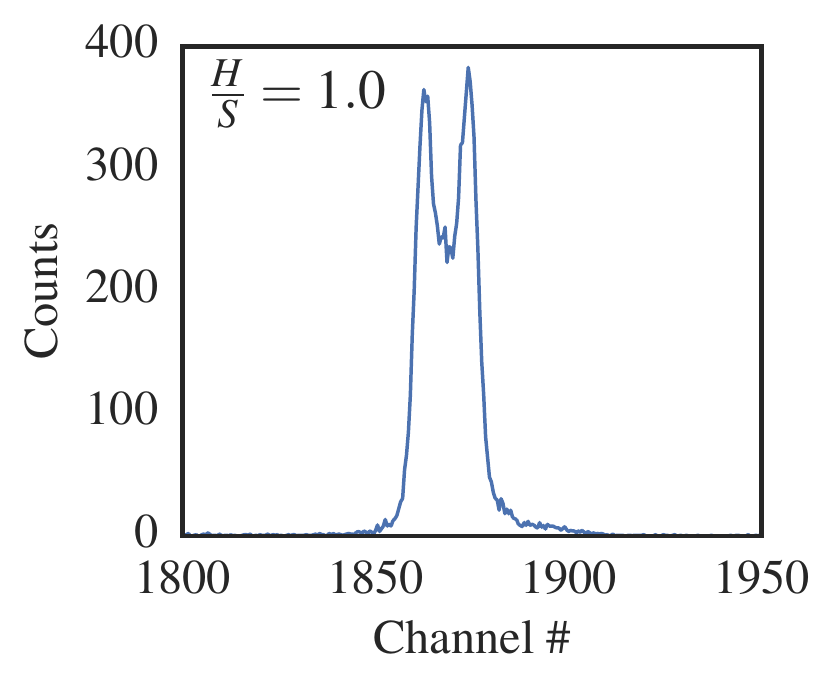}
  \includegraphics[]{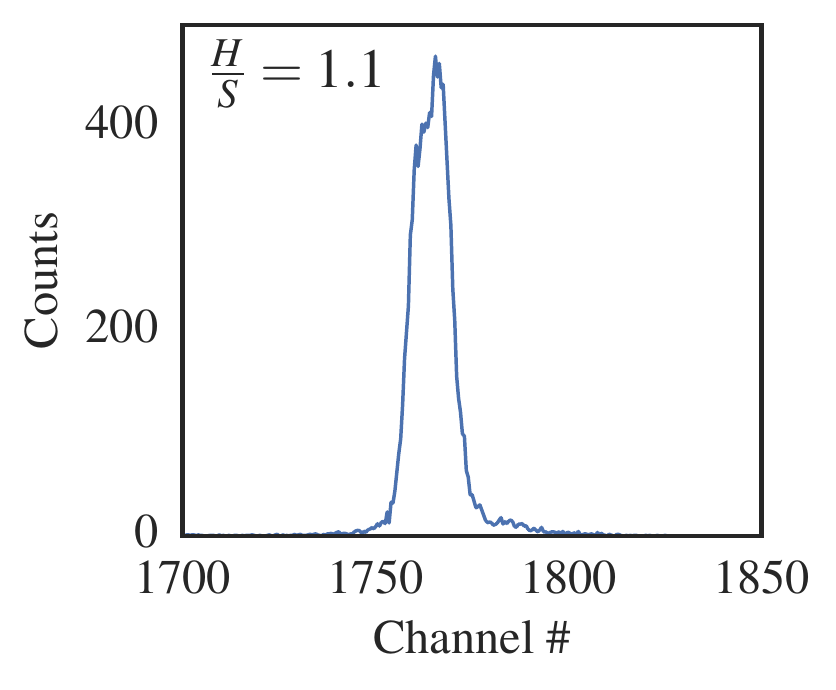}
  \caption{Using the ray tracing procedure described, an unfocused peak from the three slit aperture can be reconstructed into the single peak that would be observed at the focal plane.}
  \label{fig:recon}
\end{figure}

In practice, this correction is implemented by associating front detection signals with back signals on an event by event basis. The resulting values can then be binned and made into a new reconstructed position histogram. Fig.\ref{fig:recon} is an example of this procedure. By changing the value of $H/S$, the reconstructed focal plane can be moved back and forth virtually.      

\subsection{Effects of Gas Pressure and Voltage}

\begin{figure*}
  \centering
  \includegraphics[]{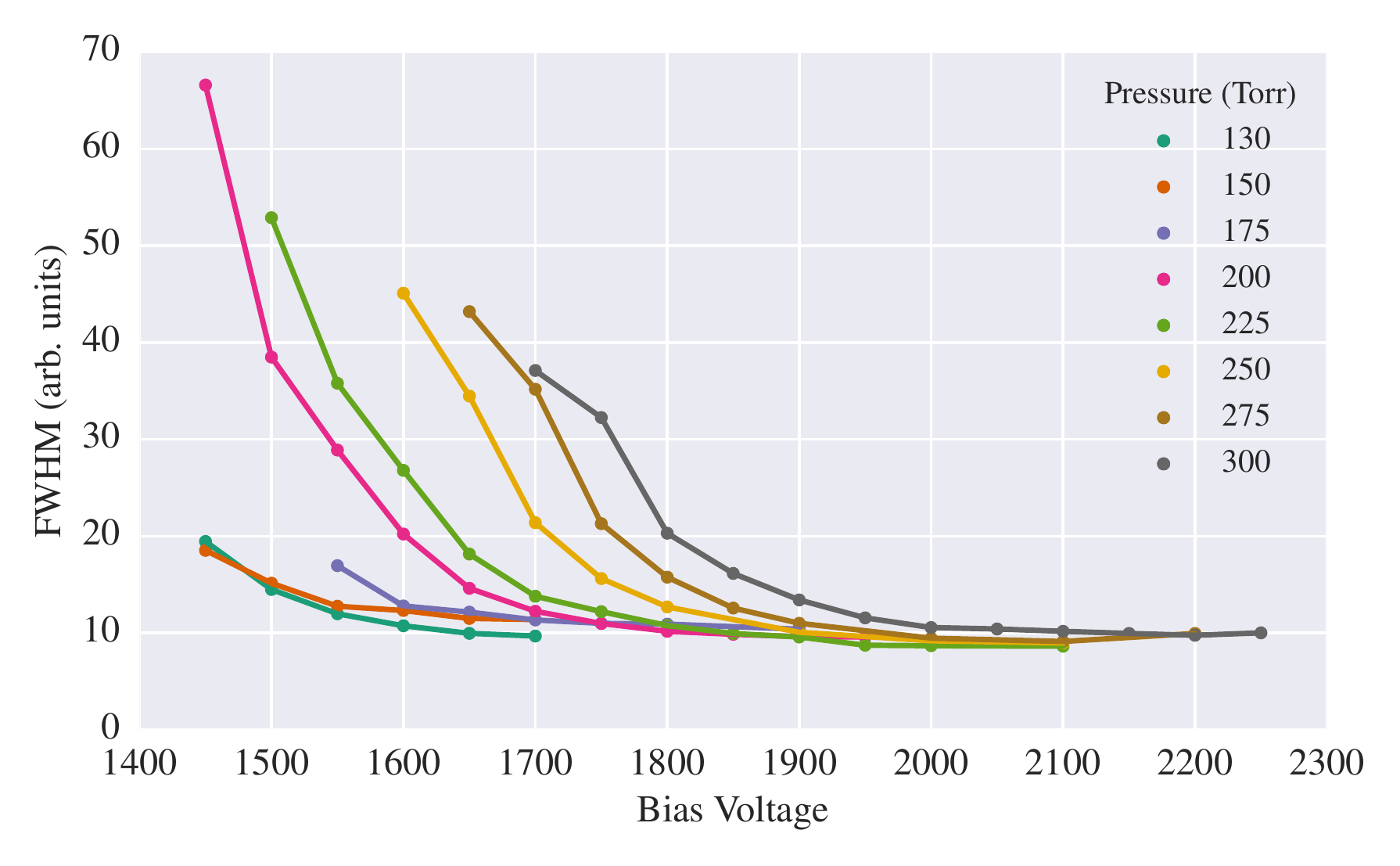}
  \caption{(color online) The FWHM of $^{12}$C$+$p elastic scattering peak at $E_{Lab} = 12$ MeV as a function of bias voltage. Different pressures are represented by different colors. The best resolution was measured at $225$ Torr and $2000\ V$}
  \label{fig:res}
\end{figure*}

$^{12}$C$(p,p)^{12}$C elastic scattering at $E_{Lab} = 12$ MeV was measured using various gas pressures and anode voltages to determine their impact on the resolution. For all of these tests, the Split-pole was positioned at $20^{\circ}$, and a small $0.25$-msr slit was used. As pointed out in \cite{xray}, low avalanche charge increases the impact of electronic noise on resolution, while higher charges start to experience photon related fluctuations. Pressures were incremented by roughly $25$ Torr from $130-300$ Torr. At each pressure, the bias was set to the lowest value that could produce a defined elastic scattering peak, typically in the $1450-1600\ V$ range, and raised until the detector experienced sparking. The results are shown in Fig.\ref{fig:res}, and they demonstrate that the best resolution occurs in the $200-250$ Torr and $1900-2100\ V$ range. Using linear fits from other spectra, which were found to have a common slope close to $0.04 \frac{\textnormal{mm}}{\textnormal{channel}}$, we estimate the optimized FWHM to be $0.35$ mm. 
 
\section{Measurement of $^{27}\!$Al$(d,p)$}

\begin{figure*}
  \centering
    \centering
    \includegraphics[]{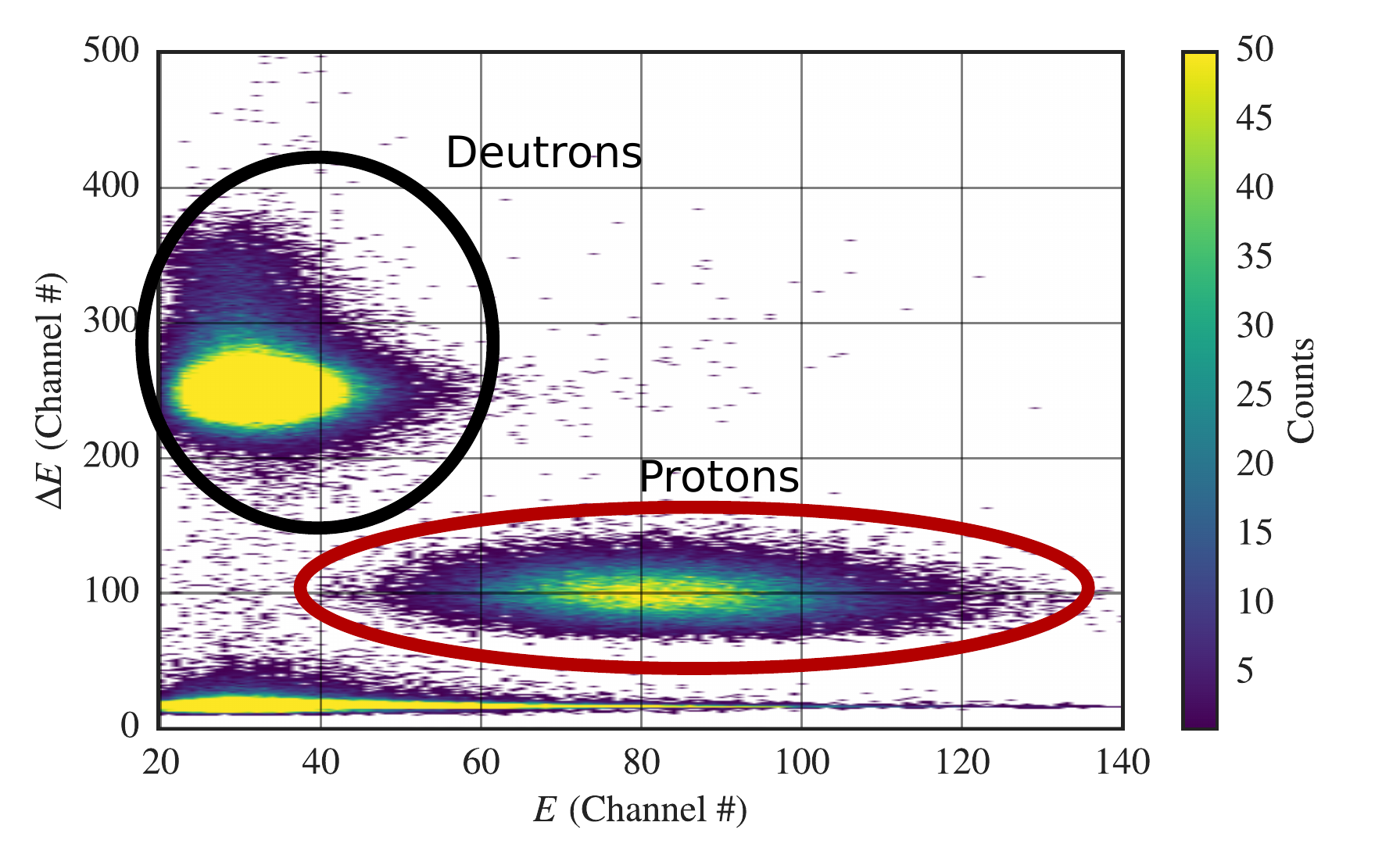}
    \caption{ (color online) $\Delta E$/E 2D spectrum from $^{27}\!$Al$(d,p)$ at $E_{Lab}=12$ MeV and $\theta_{Lab}=25^{\circ}$. The horizontal axis is the amount of energy deposited into the scintillator, while
      the vertical is the energy lost in the $\Delta E$ proportionality counter. {The two observed particle groups have been circled and labeled. The high energy tail on
      the deuteron group is due to pile up events in the detector.}}
    \label{fig:2D}
  \end{figure*}

\begin{figure*}
    \centering
    \includegraphics[]{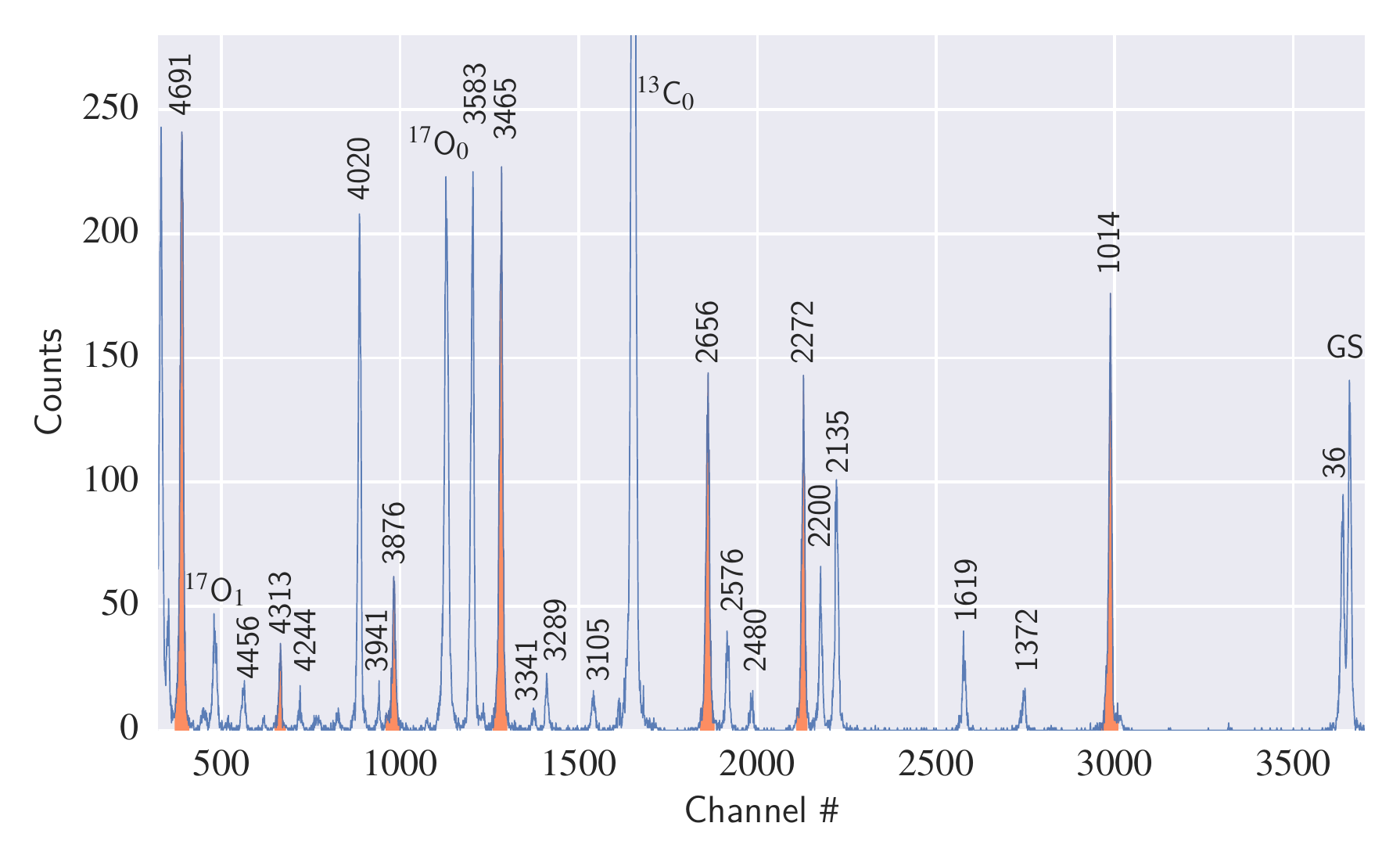}
    \caption{ (color online) Front position section gated on the proton group at $E_{Lab}=12$ MeV and $\theta_{Lab}=25^{\circ}$. Peaks used for the energy calibration are highlighted and labeled by their energies in keV.
    All other labels are the deduced energies from this work, as reported in Sec.IV.C. {Unlabeled peaks were unobserved at other angles due to lower statistics; thus, the reaction that produced them could not be identified with certainty and they are excluded from the reported energy values.}}
  \label{fig:spectra}
\end{figure*}

Detector performance was tested by analyzing reaction products of $^{27}$Al$(d,p)$. A
beam of $12$-MeV $^2$H$^{+}$ was provided by the TUNL FN Tandem Van de Graaff accelerator. A solid angle of $0.25$ msr was
chosen to minimize the effects of kinematic broadening. The detector was filled to $225$ Torr and the position sections were biased to $1800-2000$ V.
A target of ${\sim} 80$ $\mu$g/cm$^2$  $^{27}\!$Al evaporated onto a 15.2 $\mu$g/cm$^2$ $^{\textnormal{nat}}$C foil was used.
A $^{\textnormal{nat}}$C target similar to the target backing was used to identify contamination peaks arising from carbon and oxygen.
The spectrograph was positioned at three angles, $\theta_{Lab}=15^{\circ},25^{\circ},35^{\circ}$, and its field was set to $0.75$ T. Example spectra from the $\Delta E$/$E$ and the front position gated on the proton group at $25^{\circ}$ are shown in Fig. \ref{fig:2D} and Fig.\ref{fig:spectra} respectively.
% At this angle an intense deuteron group was observed. However, this group was absent at the two other angles, indicating they are most likely coming from secondary scattered beam particles.

\subsection{Energy Calibration Method}

Peaks on the focal plane represent energy levels in the residual nucleus. Using
states of known energy, focal plane spectra can be calibrated, and the remaining
energy levels can be deduced. Since the focal plane surface is curved, the relationship between $\rho$, the radius of curvature,
and ADC channel number is not linear \cite{enge}. Use of a polynomial fit corrects for curvature across
the focal plane. Once a fit is found, a value for $\rho$ can be determined for any peak in the spectrum using Eq. \ref{eq:lorentz}.
The peak centroids and energy levels used for calibration contribute to the statistical
uncertainty of the calculated energy levels. 

The process outlined above offers two distinct problems: propagation of uncertainties through the relativistic collision kinematics
and statistically rigorous regression of the polynomial fit.
For this work, uncertainty propagation was performed by using a Monte Carlo method.
This method estimates uncertainties of a target function by drawing random samples from the distributions of the dependent variables. 
For magnetic rigidities, this involves treating previous experimental values for energy levels as normal distributions. Random samples
are drawn from these distributions and then used to solve the kinematic equations.
After enough samples have been drawn, a histogram of the solutions to the kinematic equations is made, and from this information estimates of the Probability Density Function (PDF)
for $\rho$ can be made, which was found to be well described by a normal distribution.
The deduced excitation energies from the calibration use the same method, but the uncertainty contributions come from both the chosen peak centroid and the coefficients from the polynomial fit. The samples for the coefficients were drawn from a Gaussian Kernel Density Estimate (KDE) to account for the non-normality of the sampled distributions discussed below \cite{kde}.       
The resulting estimated energies, however, were found to be normally distributed.  
 
The challenge associated with the polynomial regression is ensuring that uncertainties in the peak centroids, $\mathbf{x}$, and calibration rigidities, $\boldsymbol{\rho}$,
are properly propagated through the calibration. The uncertainty should be reflected in the polynomial coefficients $(\theta_0,\ldots,\theta_N)$, where $N$
is the order of the polynomial. This problem comes down to finding the joint probability of $\boldsymbol{\theta}$ given the calibration points, $P(\boldsymbol{\theta}|\boldsymbol{\rho},\mathbf{x})$.  
For this work a Bayesian perspective and methodology was adopted.   
The joint probability can be inferred using Bayes Theorem \cite{bayes}:

\begin{equation}
  \label{eq:11}
  P(\boldsymbol{\theta}|\boldsymbol{\rho},\mathbf{x}) = \frac{P(\boldsymbol{\rho}|\mathbf{x},\boldsymbol{\theta})P(\boldsymbol{\theta})P(\mathbf{x})}{\int_{\mathbf{x}}\int_{\boldsymbol{\theta}} P(\boldsymbol{\rho}|\mathbf{x},\boldsymbol{\theta})P(\boldsymbol{\theta}) P(\mathbf{x}) d\mathbf{x} d\boldsymbol{\theta}},
\end{equation}

\noindent where $P(\boldsymbol{\theta}|\boldsymbol{\rho},\mathbf{x})$ is called the posterior, $P(\boldsymbol{\theta})$ and $P(\mathbf{x})$ are called priors, and
$P(\boldsymbol{\rho}|\mathbf{x},\boldsymbol{\theta})$ is called the likelihood. Priors are probability distributions assigned based on
knowledge about $\boldsymbol{\theta}$ and $\mathbf{x}$ before the calibration. 
On the other hand, the likelihood function gives the probability of measuring the observed data points according
to the model parameters.
The integral in the denominator of the right hand side provides an overall normalization so that the posterior is a proper PDF.
Eq. \!\ref{eq:11} gives a method to find the posterior once priors are chosen and a likelihood function for the data has been assigned.      

The polynomial calibration of the focal plane used the following model:

\begin{align}
  \label{eq:4}
  \begin{split}
    P(x_i) \sim \mathcal{N}(x_{obs_i},\,\sigma_{obs_i}^2), \\
    P(\theta_j) \sim \mathcal{N}(0,\,10^2) \\
    f(\boldsymbol{\theta},\,x_i) = \sum_{j=0}^N \theta_j x_i^j, \\
    P(\rho_i|\boldsymbol{\theta},\,x_i) \sim \mathcal{N}(f(\boldsymbol{\theta},\,x_i), \sigma_{\rho_i}^2), \\ 
  \end{split}
\end{align}

\noindent where the notation $Z \sim \mathcal{N}(\mu,\,\sigma^2)$ indicates that a random variable $Z$ is distributed according to a normal distribution with mean $\mu$ and variance $\sigma^2$. Each calibration point consists of a channel value, $x_{obs_i}$, and a radius of curvature, $\rho_i$ along with their associated variances $\sigma_{obs_i}^2$ and $\sigma_{\rho_i}^2$.
For a polynomial fit of order $N$, $\boldsymbol{\theta}$ is the vector of $N+1$ polynomial coefficients, and the polynomial function $f(\boldsymbol{\theta},\,x_i)$ is defined by $\theta_0x^0+\theta_1x^1+\ldots+\theta_Nx^N$. The choice of $\mathcal{N}(0,\,100)$ for $P(\boldsymbol{\theta})$ will be discussed more below.   
If there are $D$ measured data points, the above model defines the likelihood function:

\begin{equation}
  \label{eq:6}
  P(\boldsymbol{\rho}|\mathbf{x},\,\boldsymbol{\theta}) = \prod_i^D \exp \bigg[ \frac{1}{2} \bigg(\frac{\rho_i-f(x_i,\,\boldsymbol{\theta})}{\sigma_{\rho_i}}\bigg)^2 \bigg],
\end{equation}

\noindent where $\rho_i$ is the independently measured value for a calibration point.
To summarize this model, each calibration peak in the spectrum has a measured channel mean $x_{obs_i}$ and variance, $\sigma_{obs_i}^2$.
These values are used to assign a normal distribution to the calibration peak channel, $x_i$. This represents an informative prior for $P(x_i)$. 
Uninformative priors are selected for the polynomial coefficients. The polynomial fit function, $f$, is used as the proposed mean for the likelihood function.
The likelihood function is evaluated at the calibration points, $\rho_i$.  
Using Eq. \ref{eq:11} these distributions are used to infer the joint distribution for the values of $\boldsymbol{\theta}$ based on the data. 

Evaluation of the posterior distribution was preformed using Markov chain Monte Carlo (MCMC) \cite{mcmc}, which estimates $P(\boldsymbol{\theta}|\boldsymbol{\rho},\mathbf{x})$ by importance sampling.
Priors were chosen for each coefficient with the form of ${\sim} \mathcal{N}(0,\,100)$. In principle the intercept could be
made to be more strict since $\rho \geq 0$, but the choice in priors, provided a wide enough coverage in values, was found to have no
appreciable difference in results. The MCMC was initialized using values from a maximum likelihood estimate in order to decrease the convergence time.
The model was set up and evaluated using the PyMC package \cite{pymc}.
Typical runs draw around $2 \times 10^5$ samples after $5 \times 10^4$ initial steps
are discarded to ensure the Markov chain has time to properly converge. Thinning is also employed as needed, but convergence times can vary greatly between
different nuclei depending on how well masses and energy levels are known. Efficient sampling of the posterior was found to be greatly helped by scaling channel numbers
around their average value (i.e. for each of the N data points, $x_i$: $x_i^{\textnormal{scaled}} = x_i - \frac{1}{N}\sum_i^Nx_i $).
Sampling was also improved by scaling the magnitude of the channel numbers. For example,
if $\rho=50$ cm, then the channels would be scaled $4000 \rightarrow 40$.

There is no guarantee that a chosen set of calibration points will produce a fit that accurately predicts energies.
Frequently this problem arises from misidentifying peaks in the spectrum.
Thus, a goodness-of-fit measure is necessary to help select a valid set of calibration points.
A reduced-$\chi^2$ statistic is available in a Bayesian framework, 
but comes out of a maximum likelihood approximation, with data that has normally distributed uncertainties, and priors that are uniformly distributed \cite{bayes}. 
However, variations in the independent variables will tend to produce higher values for $\chi^2$, which could lead to the rejection of an other wise satisfactory
calibration set. In order to integrate these variations into a maximum likelihood estimate, a quantity we will call $\delta^2$ is defined as: 

\begin{multline}
  \label{eq:5}
  \delta^2 = \frac{1}{2K}\sum_{\alpha=0}^{K}\bigg[\frac{1}{N-\nu} \sum_{i=0}^{N}
  \bigg(\frac{f(x_{\alpha i};\boldsymbol{\theta})-\mu_{\rho_i}}{\sigma_{\rho_i}}\bigg)^2 \\
  + \frac{1}{M} \sum_{j=0}^{M} \bigg(\frac{x_{\alpha j}-\mu_{x_j}}{\sigma_{x_j}}\bigg)^2 \bigg]
\end{multline}

Where $N$ is the number of measured $\rho$ values, $\nu$ is the number of fitting parameters, $M$ is the number of centroids,
and $K$ is the number of centroid samples drawn. The factor of $1/2$ accounts for each term approaching unity
when the fitted parameters describe the data well. This quantity is again based on a maximum likelihood approximation applied to normally distributed
uncertainties with uniform priors, but it serves as a useful approximation for the goodness-of-fit of the $\rho$ versus channel calibration.
This method clearly distinguished misidentified peaks without giving false negatives arising from channel uncertainties.
It was found that $\delta^2 < 5$ usually indicates a fit free from misidentified states and is worth further examination.

The techniques outlined above define statistically sound procedures for uncertainty propagation and $\rho$ versus channel fitting for focal plane energy calibration.
{These procedures have the advantage of not approximating the influence of the multiple sources of uncertainty, and creating a general framework which
can be expanded as dictated by the experiment.}

\subsection{$^{28}$Al Calibration}

States from $^{28}\!$Al were identified and matched to peaks in the spectrum. 
Level energies from Ref. \cite{levels} were used both as calibration values and as comparisons for predicted energies.
Initially, a set of seven calibration states were chosen for each angle. 
A second order polynomial was chosen due to the third order term being consistent with zero.

\begin{figure*}
  \centering
  \includegraphics[width=\linewidth]{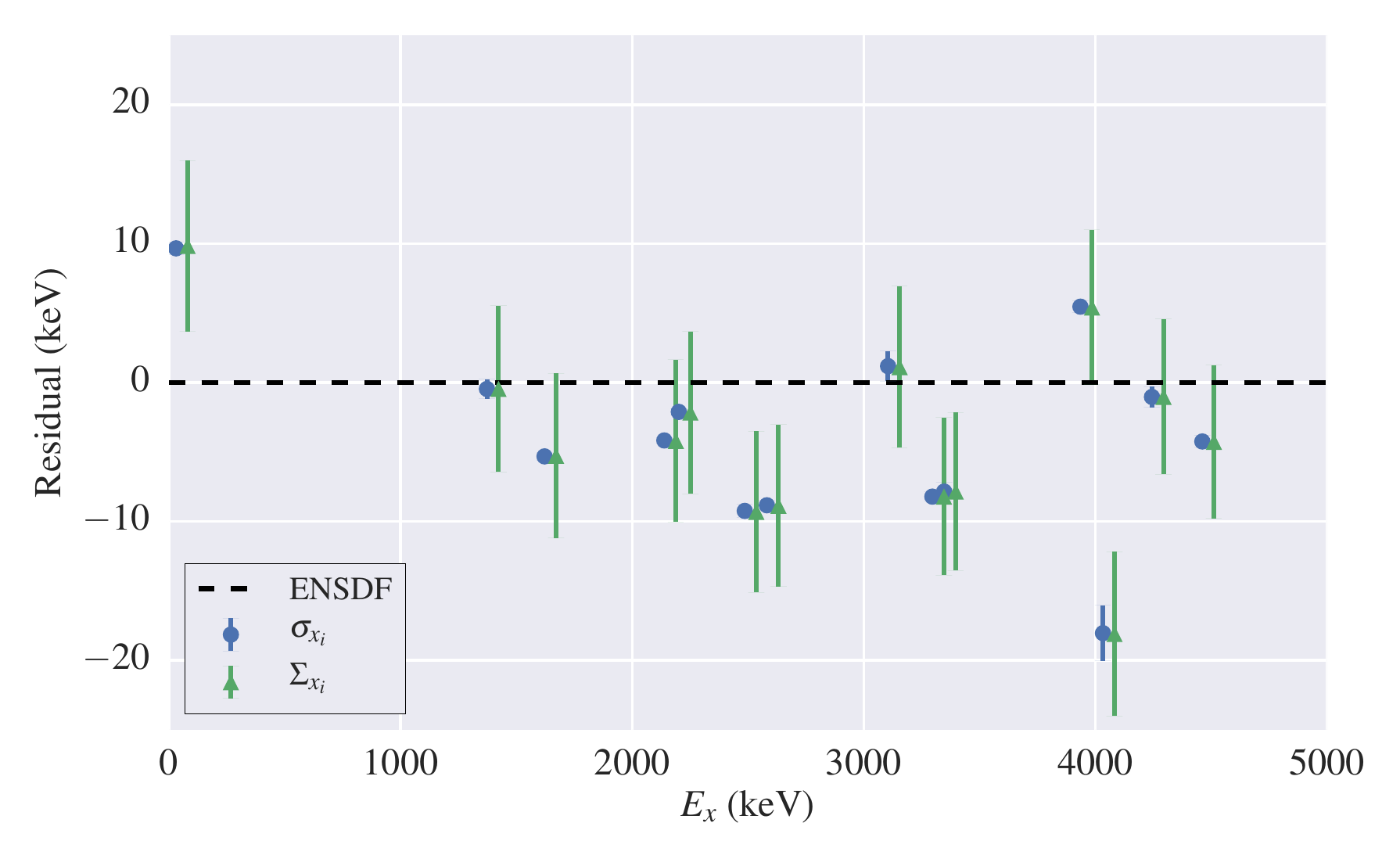}
  \caption{(color online) Residual plot of the Bayesian (blue) and {Bayesian with adjusted uncertainty (green)} calibrations for $\theta_{Lab}=25^{\circ}$. {Calibration points are not included, so only predicted energies are shown.} The excitation energies of the points {with adjusted uncertainty} have been shifted by 50 keV for visibility. The error bars represent the statistical uncertainty of the fit added in quadrature with the uncertainty reported in Ref. \protect\cite{levels}. It is clear that
    the statistical uncertainties from the fit alone are inconsistent with previously reported values. However, general agreement is found when an {additional uncertainty is fit during the calibration.}}
  \label{fig:nndc}
\end{figure*}

For the case of $^{28}$Al, most of the strongly populated states are known to sub-keV precision, which leads to small statistical uncertainties
in the fit. These small uncertainties make the deduced values inconsistent with those reported in the ENSDF database \cite{levels}.
Thus, an additional source of uncertainty is needed to account for the measured energy values.   
Sources of systematic uncertainty from experimental parameters (i.e reaction angle, beam energy, target effects)
have a minimal effect on the deduced excitation energies due to the calibration process.
For example, if the beam energy is off $\sim\! 5$ keV, then all of the calibration points, assuming they are from the
same reaction, will be shifted by roughly the same amount. Therefore, the calibration's intercept will change, and the effect is canceled out in the predicted energies.
The same arguments hold for any systematic effect that is equal for all points.  
Following these considerations, the main source of systematic uncertainty from these effects is the energy dependence of the straggling through the target.
This effect was estimated to be $0.2$ keV, which does not improve the agreement of our results with ENSDF.

The other possible sources of systematic uncertainty arise from the detector response.
In the polynomial model outlined above, the detector response is assumed to be linear; however, deviations from this assumption will cause the model to incorrectly predict
energies. In order to account for these possible effects, an extension to the Bayesian framework was used. In this model the uncertainty
for the peak centroids {(which we refer to as the "adjusted uncertainty")} is considered to be of the form:

\begin{equation}
  \label{eq:1}
  \Sigma_{x_i}^2 = \sigma_{x_i}^2+\sigma'^{2} , 
\end{equation}

where $\sigma_{x_i}$ is the observed statistical uncertainty in a given peak, $\sigma'$ is an uncertainty that is not directly measured,
and $\Sigma_{x_i}$ is the new {adjusted uncertainty} for a given peak.
The purpose of $\sigma'$ is to broaden the
normal distribution associated with each peak to a degree dictated by the available data.
This broadening accounts
for systematic effects in our position measurement, but does not assume a cause or a fixed value. Rather, it is merely another
model parameter to be estimated during calibration.  
In practice, this is done by extending the Bayesian model with a prior distribution for $\sigma'$ and using the same MCMC method to infer its value during
the polynomial regression.

The choice of prior for $\sigma'$ is more nuanced than for the other parameters previously mentioned.
Recall that our data reflects the influence of an additional uncertainty that cannot be directly estimated.
Thus, the prior must encode a source of uncertainty that is larger than the observed statistical uncertainties,
but not large enough to affect the calibration process.
These considerations lead to the adoption of a Half-Cauchy distribution for the precision, $\tau' \equiv 1/\sigma'^2$,
which is a simple transformation of the standard deviation that was found to improve MCMC convergence.   
The Half-Cauchy distribution, written $\textnormal{HalfCauchy}(\alpha,\beta)$, 
is parameterized by $\alpha$, the location parameter, and $\beta$, the scale parameter. The Half-Cauchy distribution
has been found to give good behavior close to 0, and also avoids the hard limits of the Uniform distribution \cite{gelman2006}.  
For this model $\tau' \sim \textnormal{HalfCauchy}(0,1)$ was adopted, but results showed no noticeable dependence on $\beta$.    

The comparison of the ENSDF values versus the original fit and the adjusted uncertainty fit
at $\theta_{Lab} = 25^{\circ}$ is shown in Figure \ref{fig:nndc}. Better agreement was found, indicating
the method produces reasonable estimates for the total uncertainty.

\subsection{Results}

\begin{table}[]
\centering
\setlength{\tabcolsep}{12pt}
\caption{$^{28}$Al Energy Levels (keV)}
\begin{tabular}{c c l}

  \hline
  This Work & &ENSDF \cite{levels} \\
  \hline 
  \hline
 36(5)   & & 30.6382(7)   \\
 1372(5) & & 1372.917(20)  \\
 1619(5) & & 1621.60(4)*   \\
 2135(5) & & 2138.910(10)  \\
 2200(5) & & 2201.43(3)   \\
 2480(5) & & 2486.20(6)   \\
 2576(5) & & 2581.81(22)  \\
 3105(5) & & 3105(1)      \\
 3289(5) & & 3296.34(4)   \\
 3341(5) & & 3347.19(4)   \\
 3583(5) & & 3591.457(9)  \\
 3941(5) & & 3935.603(18)  \\
 4020(5) & & 4033(3)      \\
 4244(5) & & 4244.49(10)  \\
 4456(5) & & 4461.97(10)  \\
 4510(9) & & 4516.94(18)  \\
  \hline
  \multicolumn{3}{r}{* An average of two states at 1620 and 1622.} \\

\end{tabular}

\label{tab:energies}
\end{table}

Using the Bayesian framework of Sec.IV.A with the systematic effects described in Sec.IV.B, a calibration was produced for each angle.
The same calibration peaks were used for each angle, and they represent strongly populated states that are well-resolved in the spectra.
The location of these states at $\theta_{Lab}=25^{\circ}$ are shown in Fig.\ref{fig:spectra}

The reported values listed in Table \ref{tab:energies} are weighted averages of the energies over all angles, with
the requirement that any candidate state be observed at more than one angle. A total of 16 states (excluding the 7 calibration states) were measured in this way.

Finally, an estimation for the detector resolution was found from the slope of the $\rho$ calibration.
We found a slope of $0.036 \frac{\textnormal{mm}}{\textnormal{channel}}$, and typical
FWHM in the spectrum of $10$-$20$ channels. These values give resolutions between $0.36$-$0.72$ mm.
The separation on the ground and first excited state implies an energy resolution of $\sim\!15$ keV.
 
\section{Conclusion}

In the present paper we have presented a description of the focal-plane detector for the Enge Split-pole spectrograph at TUNL. The methods used to fabricate the detector emphasize ease of maintenance, flexibility for future system upgrades, and the resolution and particle identification necessary to study nuclear reactions of interest for astrophysics. Geant4 simulations found that
our design decision to put the position sensitive cathodes in the particle path had a negligible effect on resolution, with most secondary scattering occurring at the entrance window. Kinematic effects on resolution were discussed, and methods for correcting them were presented. These methods were: an empirical fit between the kinematic factor and detector position and using the detector's two position sections for ray tracing. Optimal gas pressure and anode bias voltages were found from $^{12}$C$(p,p)$ scattering. A measurement of the $^{27}\!$Al$(d,p)$ reaction was performed, and confirmed the design provides sub-millimeter resolution. A Bayesian method for energy calibration was also discussed, which ensures that uncertainties in both calibration peaks and energy levels are accounted for. This method was then extended to include considerations of systematic effects in the detector, and used to extract energy levels of $^{28}\!$Al. These values were reported and compared to previous measurements, and are in excellent agreement for most cases.    

\section{Acknowledgments}

The authors would like to thank the technical staff at TUNL for their contributions and assistance.  
This material is based upon work supported by the U.S. Department of Energy, Office of Science, Office of Nuclear Physics, under Award Number DE-SC0017799 and under Contract No. DE-FG02-97ER41041.

% do the biliography:
\bibliographystyle{IEEEbib}
\bibliography{detector}

\begin{thebibliography}{10}

\bibitem{starstuff}
George Wallerstein, Icko Iben, Peter Parker, Ann~Merchant Boesgaard, Gerald~M.
  Hale, Arthur~E. Champagne, Charles~A. Barnes, Franz Kappeler, Verne~V. Smith,
  Robert~D. Hoffman, Frank~X. Timmes, Chris Sneden, Richard~N. Boyd, Bradley~S.
  Meyer, and David~L. Lambert,
\newblock ``Synthesis of the elements in stars: forty years of progress,''
\newblock {\em Rev. Mod. Phys.}, vol. 69, pp. 995--1084, Oct 1997.

\bibitem{iliadis}
Christian Iliadis,
\newblock {\em Nuclear Physics of Stars},
\newblock Wiley-VCH Verlag GmbH \& Co. KGaA, 2015.

\bibitem{stable}
D~W Bardayan,
\newblock ``Transfer reactions in nuclear astrophysics,''
\newblock {\em Journal of Physics G: Nuclear and Particle Physics}, vol. 43,
  no. 4, pp. 043001, 2016.

\bibitem{RIB}
Karlheinz Langanke and Hendrik Schatz,
\newblock ``The role of radioactive ion beams in nuclear astrophysics,''
\newblock {\em Physica Scripta}, vol. 2013, no. T152, pp. 014011, 2013.

\bibitem{enge}
Harald~A. Enge,
\newblock ``Magnetic spectrographs for nuclear reaction studies,''
\newblock {\em Nuclear Instruments and Methods}, vol. 162, no. 1, pp. 161 --
  180, 1979.

\bibitem{splitpole}
J.E. Spencer and H.A. Enge,
\newblock ``Split-pole magnetic spectrograph for precision nuclear
  spectroscopy,''
\newblock {\em Nuclear Instruments and Methods}, vol. 49, no. 2, pp. 181 --
  193, 1967.

\bibitem{cosmos}
K.~Setoodehnia, R.~Longland, et~al.,
\newblock ``Proceedings of xiv nuclei in the cosmos conference,''
\newblock {\em JPS Conf. Proc.}, 2016.

\bibitem{hale}
Stephen E.~Hale Jr.,
\newblock {\em $^{22} \mathrm{Ne}(p,\gamma)^{23}\mathrm{Na},
  ^{23}\mathrm{Na}(p,\gamma)^{24}\mathrm{Mg}$ Globular Clustar Abundance
  Anomalies},
\newblock Ph.D. thesis, University of North Carolina at Chapel Hill, 1999.

\bibitem{knoll}
Glenn~F Knoll,
\newblock {\em {Radiation detection and measurement; 4th ed.}},
\newblock Wiley, New York, NY, 2010.

\bibitem{etch}
C.L Morris, L.G Atencio, W.E Sondheim, S.J Seestrom, M.W Rawool-Sullivan, P.L
  McGaughey, D.M Lee, W.W Kinnison, M.L Brooks, and V~Armijo,
\newblock ``Electric discharge etching of thin metalized plastic films,''
\newblock {\em Nuclear Instruments and Methods in Physics Research Section A:
  Accelerators, Spectrometers, Detectors and Associated Equipment}, vol. 379,
  no. 2, pp. 243 -- 246, 1996.

\bibitem{msu}
R.G. Markham and R.G.H. Robertson,
\newblock ``High resolution position-sensitive proportional counter,''
\newblock {\em Nuclear Instruments and Methods}, vol. 129, no. 1, pp. 131 --
  140, 1975.

\bibitem{heavy}
F.L.H. Wolfs, D.C. Bryan, K.L. Kurz, D.M. Herrick, P.A.A. Perera, and C.A.
  White,
\newblock ``Segmented focal plane detector for light and heavy ions,''
\newblock {\em Nuclear Instruments and Methods in Physics Research Section A:
  Accelerators, Spectrometers, Detectors and Associated Equipment}, vol. 317,
  no. 1, pp. 221 -- 234, 1992.

\bibitem{parikh}
Anuj~Ramesh Parikh,
\newblock {\em Production of $^{26}\mathrm{Al}$ in Oxygen-Neon-Magnesium
  Novae},
\newblock Ph.D. thesis, Yale University, 2006.

\bibitem{argonne_det}
J.R. Erskine, T.H. Braid, and J.C. Stoltzfus,
\newblock ``An ionization-chamber type of focal-plane detector for heavy
  ions,''
\newblock {\em Nuclear Instruments and Methods}, vol. 135, no. 1, pp. 67 -- 82,
  1976.

\bibitem{vert}
H.~Lindner, H.~Angerer, and G.~Hlawatsch,
\newblock ``A new multiwire proportional chamber with fast single strip readout
  and individual analog to digital conversion,''
\newblock {\em Nuclear Instruments and Methods in Physics Research Section A:
  Accelerators, Spectrometers, Detectors and Associated Equipment}, vol. 273,
  no. 1, pp. 444 -- 446, 1988.

\bibitem{BNLDetector}
E.~Beardsworth, J.~Fischer, S.~Iwata, M.J. Levine, V.~Radeka, and C.E. Thorn,
\newblock ``Multiwire proportional chamber focal-plane detector,''
\newblock {\em Nuclear Instruments and Methods}, vol. 127, no. 1, pp. 29 -- 39,
  1975.

\bibitem{chips}
Data Delay Devices,
\newblock {\em 10-TAP SIP DELAY LINE T$_D$/T$_R$=5 (SERIES 1507)}, 2014.

\bibitem{widths}
G.~C. Smith, J.~Fischer, and V.~Radeka,
\newblock ``Capacitive charge division in centroid finding cathode readouts in
  mwpcs,''
\newblock {\em IEEE Transactions on Nuclear Science}, vol. 35, no. 1, pp.
  409--413, Feb 1988.

\bibitem{gas}
T.R. Ophel, L.K. Fifield, W.N. Catford, N.A. Orr, C.L. Woods, A.~Harding, and
  G.P. Clarkson,
\newblock ``The identification and rejection of energy-degraded events in gas
  ionization counters,''
\newblock {\em Nuclear Instruments and Methods in Physics Research Section A:
  Accelerators, Spectrometers, Detectors and Associated Equipment}, vol. 272,
  no. 3, pp. 734 -- 749, 1988.

\bibitem{cremat}
Cremat,
\newblock {\em CR-110 charge sensitive preamplifier: application guide}, 2014.

\bibitem{gobain}
Saint-Gobain Crystals,
\newblock {\em BC-400,BC-404,BC-408,BC-412,BC-416 Data Sheet}, 8 2016.

\bibitem{wrapping}
R.~Wojcik, B.~Kross, S.~Majewski, A.G. Weisenberger, and C.~Zorn,
\newblock ``Embedded waveshifting fiber readout of long scintillators,''
\newblock {\em Nuclear Instruments and Methods in Physics Research Section A:
  Accelerators, Spectrometers, Detectors and Associated Equipment}, vol. 342,
  no. 2, pp. 416 -- 435, 1994.

\bibitem{fibers}
Saint-Gobain Crystals,
\newblock {\em Scintillating Optical Fibers}, 8 2016.

\bibitem{hamamatsu}
Hamamatsu,
\newblock {\em Photomultiplier tube assembly H6524 Specifications}, 2016.

\bibitem{Geant4}
S.~Agostinelli, J.~Allison, K.~Amako, J.~Apostolakis, H.~Araujo, P.~Arce,
  M.~Asai, D.~Axen, S.~Banerjee, G.~Barrand, F.~Behner, L.~Bellagamba andJ.
  Boudreau, L.~Broglia, A.~Brunengo, H.~Burkhardt, S.~Chauvie, J.~Chuma,
  R.~Chytracek, G.~Cooperman, G.~Cosmo, P.~Degtyarenko, A.~Dell'Acqua,
  G.~Depaola, D.~Dietrich, R.~Enami, A.~Feliciello, C.~Ferguson, H.~Fesefeldt,
  G.~Folger, F.~Foppiano, A.~Forti, S.~Garelli, S.~Giani, R.~Giannitrapani,
  D.~Gibin, J.J.~Gomez Cadenas, I.~Gonzalez, G.~Gracia Abril, G.~Greeniaus,
  W.~Greiner, V.~Grichine, A.~Grossheim, S.~Guatelli, P.~Gumplinger,
  R.~Hamatsu, K.~Hashimoto, H.~Hasui, A.~Heikkinen, A.~Howard, V.~Ivanchenko,
  A.~Johnson, F.W. Jones, J.~Kallenbach, N.~Kanaya, M.~Kawabata, Y.~Kawabata,
  M.~Kawaguti, S.~Kelner, P.~Kent, A.~Kimura, T.~Kodama, R.~Kokoulin,
  M.~Kossov, H.~Kurashige, E.~Lamanna, T.~Lampen, V.~Lara, V.~Lefebure, F.~Lei,
  M.~Liendl, W.~Lockman, F.~Longo, S.~Magni, M.~Maire, E.~Medernach,
  K.~Minamimoto, P.~Mora de~Freitas, Y.~Morita, K.~Murakami, M.~Nagamatu,
  R.~Nartallo, P.~Nieminen, T.~Nishimura, K.~Ohtsubo, M.~Okamura, S.~O'Neale,
  Y.~Oohata, K.~Paech, J.~Perl, A.~Pfeiffer, M.G. Pia, F.~Ranjard, A.~Rybin,
  S.~Sadilov, E.~Di Salvo, G.~Santin, T.~Sasaki, N.~Savvas, Y.~Sawada,
  S.~Scherer, S.~Sei, V.~Sirotenko, D.~Smith, N.~Starkov, H.~Stoecker,
  J.~Sulkimo, M.~Takahata, S.~Tanaka, E.~Tcherniaev, E.~Safai Tehrani,
  M.~Tropeano, P.~Truscott, H.~Uno, L.~Urban, P.~Urban, M.~Verderi, A.~Walkden,
  W.~Wander, H.~Weber, J.P. Wellisch, T.~Wenaus, D.C. Williams, D.~Wright,
  T.~Yamada, H.~Yoshida, and D.~Zschiesche,
\newblock ``Geant4-a simulation toolkit,''
\newblock {\em Nuclear Instruments and Methods in Physics Research Section A:
  Accelerators, Spectrometers, Detectors and Associated Equipment}, vol. 506,
  no. 3, pp. 250--303, 2003.

\bibitem{ROOT}
Rene Brun and Fons Rademakers,
\newblock ``Root - an object oriented data analysis framework,''
\newblock {\em Nuclear Instruments and Methods in Physics Research Section A:
  Accelerators, Spectrometers, Detectors and Associated Equipment}, vol. 389,
  no. 1, pp. 81--86, 1997.

\bibitem{ionoptics}
M.N. Viswesvariah and N.~Sarma,
\newblock ``The ion-optics of a split pole magnetic spectrograph,''
\newblock {\em Nuclear Instruments and Methods}, vol. 54, no. 2, pp. 181 --
  189, 1967.

\bibitem{matrix}
A.~Cunsolo, F.~Cappuzzello, A.~Foti, A.~Lazzaro, A.L. Melita, C.~Nociforo,
  V.~Shchepunov, and J.S. Winfield,
\newblock ``Ion optics for large-acceptance magnetic spectrometers: application
  to the magnex spectrometer,''
\newblock {\em Nuclear Instruments and Methods in Physics Research Section A:
  Accelerators, Spectrometers, Detectors and Associated Equipment}, vol. 484,
  no. 1, pp. 56 -- 83, 2002.

\bibitem{partA}
D.L. Hendrie,
\newblock ``Iii.c - magnetic detection of charged particles,''
\newblock in {\em Nuclear Spectroscopy and Reactions, Part A}, JOSEPH CERNY,
  Ed., vol. 40, Part A of {\em Pure and Applied Physics}, pp. 365 -- 412.
  Elsevier, 1974.

\bibitem{raytrace}
D.~Shapira, R.M. Devries, H.W. Fulbright, J.~Tōke, and M.R. Clover,
\newblock ``The rochester heavy ion detector,''
\newblock {\em Nuclear Instruments and Methods}, vol. 129, no. 1, pp. 123 --
  130, 1975.

\bibitem{ugalde}
C.~Ugalde, A.~E. Champagne, S.~Daigle, C.~Iliadis, R.~Longland, J.~R. Newton,
  E.~Osenbaugh-Stewart, J.~A. Clark, C.~Deibel, A.~Parikh, P.~D. Parker, and
  C.~Wrede,
\newblock ``Experimental evidence for a natural parity state in
  $^{26}\mathrm{Mg}$ and its impact on the production of neutrons for the $s$
  process,''
\newblock {\em Phys. Rev. C}, vol. 76, pp. 025802, Aug 2007.

\bibitem{xray}
J.~Fischer, V.~Radeka, and G.C. Smith,
\newblock ``X-ray position detection in the region of 6 μm rms with wire
  proportional chambers,''
\newblock {\em Nuclear Instruments and Methods in Physics Research Section A:
  Accelerators, Spectrometers, Detectors and Associated Equipment}, vol. 252,
  no. 2, pp. 239 -- 245, 1986.

\bibitem{kde}
Alan~Julian Izenman,
\newblock ``Recent developments in nonparametric density estimation,''
\newblock {\em Journal of the American Statistical Association}, vol. 86, no.
  413, pp. 205--224, 1991.

\bibitem{bayes}
D~S~Sivia and J~Skilling,
\newblock {\em Data Analysis A Bayesian Tutorial},
\newblock 01 2006.

\bibitem{mcmc}
C.~Andrieu, A.~Doucet, and C.~P. Robert,
\newblock ``Computational advances for and from bayesian analysis,''
\newblock {\em Statistical Science}, vol. 19, no. 1, pp. 118--127, 2013.

\bibitem{pymc}
Chris Fonnesbeck, Anand Patil, David Huard, and John Salvatier,
\newblock ``Pymc 2.3.6,'' https://github.com/pymc-devs/pymc.

\bibitem{levels}
M.~Shamsuzzoha Basunia,
\newblock ``Nuclear data sheets for a = 28,''
\newblock {\em Nuclear Data Sheets}, vol. 114, no. 10, pp. 1189 -- 1291, 2013.

\bibitem{gelman2006}
Andrew Gelman,
\newblock ``Prior distributions for variance parameters in hierarchical models
  (comment on article by browne and draper),''
\newblock {\em Bayesian Analysis}, vol. 1, no. 3, pp. 515--534, 09 2006.

\end{thebibliography}

\end{document}